\documentclass[9pt, final,twocolumn]{IEEEtran}
\usepackage{graphicx}
\usepackage{amsthm}
\usepackage{epsfig}
\usepackage{latexsym}
\usepackage{amsfonts}
\usepackage{here}
\usepackage{rawfonts}
\usepackage[latin1]{inputenc}
\usepackage[T1]{fontenc}
\usepackage{calc}
\usepackage{url}
\usepackage{enumerate}
\usepackage{color}
\usepackage[tbtags]{amsmath}
\usepackage{amssymb}
\usepackage{amsmath}
\usepackage{upref}
\usepackage{epic,eepic}
\usepackage{times}
\usepackage{dsfont}
\usepackage{comment}
\usepackage{cite}
\usepackage{bbm} 
\usepackage{multirow}

\usepackage{threeparttable}

\renewcommand{\Pr}{\ensuremath{\operatorname{Pr}}}

\newif\ifrevision
\newcommand{\highlight}[1]{%
\ifrevision
\textcolor{red}{#1}%
\else
#1%
\fi
}

\newif\ifrevisions
\newcommand{\highlights}[1]{%
\ifrevisions
\textcolor{blue}{#1}%
\else
#1%
\fi
}

\newtheorem{theorem}{\bf Theorem}

\newtheorem{definition}{\bf Definition}

\usepackage{dsfont}

\newlength{\aligntop}
\newlength{\alignbot}
 

\IEEEoverridecommandlockouts

\graphicspath{{./Figures/}}

\usepackage{algorithm}
\usepackage{algorithmic}

\makeatletter
\newcommand\semihuge{\@setfontsize\semihuge{19.3}{25}}
\makeatother

\makeatletter
\newcommand\semismall{\@setfontsize\semihuge{12.4}{15}}
\makeatother

\usepackage{subfigure}

\usepackage{tablefootnote}

\begin{document}

\revisionfalse
\revisionsfalse

\title{Joint Source-Channel Coding: Fundamentals and Recent Progress in Practical Designs \vspace*{-0em}}

\author{Deniz G\"und\"uz,~\IEEEmembership{Fellow,~IEEE}, Mich\`{e}le A. Wigger,~\IEEEmembership{Senior Member,~IEEE}, Tze-Yang Tung~\IEEEmembership{Member,~IEEE},\\ Ping Zhang,~\IEEEmembership{Fellow,~IEEE}, Yong Xiao,~\IEEEmembership{Senior Member,~IEEE},
\thanks{D. G\"und\"uz and T.-Y. Tung are with the Department of Electrical and Electronic Engineering, Imperial College London, SW7 2BT, London, U.K. (e-mail: \{d.gunduz, tze-yang.tung14\}@imperial.ac.uk). M. A. Wigger is with LTCI, Telecom Paris, IP Paris, 91120 Palaiseau, France (e-mail: michele.wigger@telecom-paris.fr). P. Zhang is with the State Key Laboratory of Networking and Switching Technology, Beijing University of Posts and Telecommunications, Beijing 100876, China (email: pzhang@bupt.edu.cn).
Y. Xiao is with the School of Electronic Information and Communications at the Huazhong University of Science and Technology, Wuhan 430074, China, also with the Peng Cheng Laboratory, Shenzhen, Guangdong 518055, China, and also with the Pazhou Laboratory (Huangpu), Guangzhou, Guangdong 510555, China (e-mail: yongxiao@hust.edu.cn).
}
\thanks{This work received funding from the UKRI for the projects AI-R (ERC Consolidator Grant, EP/X030806/1), SONATA (EPSRC-EP/W035960/1), the SNS JU project 6G-GOALS under the EU's Horizon program (Grant Agreement No. 101139232), and the National Natural Science Foundation of China (Grant No. 62293481 and 62071193).}
}

\maketitle
%

\begin{abstract}
Semantic- and task-oriented communication has emerged as a promising approach to reducing the latency and bandwidth requirements of next-generation mobile networks by transmitting only the most relevant information needed to complete a specific task at the receiver. This is particularly advantageous for machine-oriented communication of high data rate content, such as images and videos, where the goal is rapid and accurate inference, rather than perfect signal reconstruction. \highlights{While semantic- and task-oriented compression can be implemented in conventional communication systems, 
joint source-channel coding (JSCC) offers an alternative end-to-end approach by optimizing compression and channel coding together, or even directly mapping the source signal to the modulated waveform. 
Although all digital communication systems today rely on separation, thanks to its modularity, JSCC is known to achieve higher performance in finite blocklength scenarios, and to avoid \textit{cliff} and the \textit{levelling-off effects} in time-varying channel scenarios.} This article provides an overview of the information theoretic foundations of JSCC, surveys practical JSCC designs over the decades, and discusses the reasons for their limited adoption in practical systems. We then examine the recent resurgence of JSCC, driven by the integration of deep learning techniques, particularly through DeepJSCC, highlighting its many surprising advantages in various scenarios. Finally, we discuss why it may be time to reconsider today's strictly separate architectures, and reintroduce JSCC to enable high-fidelity, low-latency communications in critical applications such as autonomous driving, drone surveillance, or wearable systems.
\end{abstract}

\section{Introduction}

\begin{figure}
		\centering
		\includegraphics[width=3.5in]{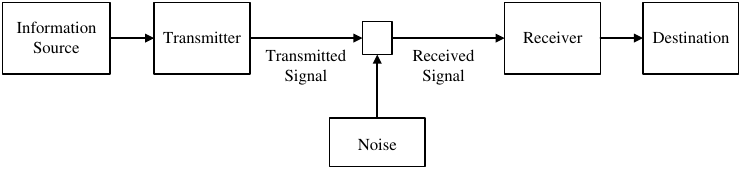}
		\caption{Schematic diagram of a general communication system according to Shannon (reproduced from \cite{Shannon}).}
  \label{f:Shannon_model}
	\end{figure}

Shannon defined the fundamental problem of communication as that of ``reproducing at one point either exactly or approximately a message selected at another point'' \cite{Shannon}. 
A generic point-to-point communication system,  reproduced in Fig. \ref{f:Shannon_model} from \cite{Shannon}, consists of an information source, a transmitter, a communication channel, and a receiver. The information source, which generates the messages to be transmitted, and the channel are typically assumed to be given, while the goal of the system designer is to come up with the transmitter and receiver mappings. The former, called the \textit{encoding function}, maps the message to a form that is amenable to be transmitted over the channel, while the latter, called the \textit{decoding function}, is a complementary mapping that recovers the message under a prescribed fidelity measure. The fidelity measure, also assumed to be given, maps every message and reconstruction pair to a real number, quantifying the goodness of the reconstruction.    

The problem stated in this form is fairly general: the information source can be a sequence of discrete symbols from a prescribed alphabet, or a function of time and other variables. In general, the information source is a stochastic process, and the communication system should operate for any source realization as it is not known at the time of design. The communication channel is also modelled as a stochastic kernel from the input alphabet space to the output alphabet space, which depend on the communication medium under consideration. We may assume the statistical properties of the information source and the channel as fixed and known, or they may belong to a certain set of possible statistics \cite{Csiszar:TIT:88, Feder:ISIT:95}, or we can also assume that the statistics are fixed yet unknown, but we can sample from these distributions as desired.

The design of a communication system in this full generality is highly challenging. The transmitter and receiver must be designed jointly considering both the source and channel characteristics, that is, their alphabets as well as statistics. Some of the earliest practical telecommunication systems, amplitude/frequency modulation (AM/FM), analog TV broadcasting, and the first generation of the mobile wireless networks have all been based on this approach. On the other hand, a fundamental result in Shannon's information theory is the Separation Theorem \cite{Shannon}, which states that for a large class of point-to-point communication scenarios characterized by large block lengths, optimal performance can be effectively approached by independently formulating the compression and error correction schemes. Equivalently, the problem of mapping the input source message to the channel input, and recovering the input signal from the noisy received channel output can be reformulated into two subproblems of \textit{source coding} and \textit{channel coding} without loss of optimality. 

The source coding problem deals with mapping the message to the minimum number of bits so that it can be reconstructed within the desired fidelity. Shannon's Source Coding Theorem identifies the fundamental limit of this problem, characterized by the so-called \textit{rate-distortion function}, considering an arbitrarily long sequence of source samples and an additive distortion measure. This function provides a bound on the minimum number of bits per source sample that must be transmitted to the receiver to achieve the desired fidelity. On the other hand, Shannon's Channel Coding Theorem identifies the fundamental limit of communications over a noisy channel, characterized by the channel capacity. Similarly to the source coding theory, channel coding theory assumes infinitely many uses of the channel, and identifies maximum number of bits per channel use that can be conveyed reliably over the channel. 

In the point-to-point setting, the Separation Theorem suggests that if the compression rate corresponding to the target distortion of the given source distribution is below the channel capacity of the link, then the desired distortion can be achieved. Conversely, if this rate is above the capacity, there is no coding scheme that can achieve this distortion level over this channel (more rigorous definitions and statements will be provided later in the paper). Thanks to the converse theorem, we conclude that the desired distortion level can either be achieved by designing separate source and channel codes, or it is impossible to achieve. 

In separate source and channel coding (SSCC), the compression scheme is designed oblivious to the channel statistics, while the channel code is designed ignoring the source characteristics. This separate design principle leads to a natural modular system architecture, allowing a single network infrastructure that can serve a great variety of services, e.g., voice, data, multimedia. Indeed, almost all contemporary communication systems adhere to such a structured layered architecture, in which compression-related functions are primarily dealt with at the application layer, at the top of the protocol stack, while channel coding functions are taken care of at the link and physical layers, at the bottom of the stack. 

This modularity in design combined with the theoretical optimality led to a natural separation among the research communities as well. Researchers in the data compression community have mainly focused on developing high performance compression algorithms by exploiting properties of particular types of source signals, leading to various standards for text, image, audio and video compression, among others. In parallel, communication researchers have focused on designing advanced coding and modulation techniques with the goal of turning the noisy wireless medium into a network of noise-free bit-pipes.  


Although a segregated architecture offers indisputable advantages in facilitating modular system design and enabling the convergence of diverse services on a shared data network infrastructure, there are instances where a joint source-channel coding (JSCC) approach becomes essential. This need arises both in various multi-terminal settings where the separated approach proves suboptimal, as well as in standard point-to-point channels. Even in cases where SSCC is asymptotically optimal, the application of independently designed source and channel codes may result in subpar performance under practical conditions characterized by finite block length, low-complexity encoding/decoding, or communication over uncertain channels.

JSCC has been a topic of continuous research since Shannon; however, there has been a significant surge in interest in recent years. This can be attributed to two main reasons, both are connected to the recent advances in machine learning (ML) technologies. Firstly, as ML tools and applications become increasingly widespread, they are expected to be deployed on mobile edge devices, requiring distributed implementations of training as well as inference tasks \cite{Gunduz:CM:20, Mao:CST:17, Imteaj:IoTJ:22, xiao2020selflearning, Li:Network:18, Chen:JSAC:21, xiao2022imitation}. These applications typically have much more strict latency requirements compared to content delivery applications, for which the current network architectures and communication schemes have been designed. The gap between the optimal performance that can be achieved through JSCC and what can be achieved by conventional SSCC tends to increase as the blocklengths become shorter. Moreover, for shorter blocklengths, resources used for channel estimation may become more significant, particularly for highly mobile environments. The modularity that separation provides is extremely valuable for larger network architectures, such as cellular networks; however, modularity can be sacrificed in point-to-point communication scenarios, where the performance gains become critical. For example, for the delivery of drone video footage to a ground station, the wireless connection of a augmented reality (AR)/virtual reality (VR) headset to a computing unit, or the exchange of LIDAR or video data among autonomous vehicles for collaborative perception, it may be crucial to go beyond the constraints of a strictly separate design.

Secondly, despite many years of efforts, the research on practical code design for JSCC has not produced codes that can be applied to a large variety of source-channel pairs with a performance-complexity trade-off that is better than, or on par with the state-of-the-art separate baselines. This has changed recently with the advances in deep learning aided JSCC design, called DeepJSCC \cite{Eirina:TCCN:19, Xu:ComMag:23}. Such codes can be easily designed for different sources modalities and datasets targeting any desired performance measure \cite{Weng:JSAC:21, Peng:GLOBECOM:2022, Xie:TSP:21, Jankowski:ISIT:22, Han:JSAC:23} providing highly competitive performance results. 

\highlight{Developments in JSCC design have been further encouraged with the recent growing interest in semantic and goal-oriented communication systems \cite{Gunduz:JSAC:23, kalfa2021towards, XY2021SemanticCommMagazine, Uysal:Network:22, Tung:JSAC:21, shao2023theory, zhang2022deep, xiao2023reasoning, XY2022ITWSSC}, driven by AR/VR applications, and the emerging concepts of metaverse \cite{Schwenteck:NL:23}, tactile/haptic communications \cite{Fettweis:BITS:21}, the Internet of senses \cite{Joda:Network:23}, and holographic communications \cite{Tataria:PIEEE:21}, that are foreseen to be the main drivers of 6G and beyond mobile networks. The motivation for semantic and goal-oriented communication comes from taming the growing network data traffic, particularly due to the emerging artificial intelligence (AI) applications involving AR/VR, autonomous vehicles and machine vision. The efforts on the communication network design side have mainly focused on increasing the capacity of the network; however, gains from physical and network layer modifications (e.g., introduction of massive number of antennas or terminals) have become increasingly costly, and incremental as we approach the corresponding fundamental limits. Hence, a complementary direction is to reduce the amount of information that must be communicated over the network. In a nutshell, the core idea behind semantic and goal-oriented communication is to transmit only the relevant information for the underlying task at the receiver, which can be considered as a form of compression \cite{Gunduz:JSAC:23}. Naturally, such an approach can be carried out purely on the application layer by employing the appropriate compression algorithm for the desired goal, reducing the number of bits that need to be conveyed. One can argue that this is already done for most multimedia applications that rely on various compression algorithms, although new types of compression algorithms need to be designed for specific inference tasks (e.g., image retrieval \cite{Jankowski:JSAC:21}, or other inference tasks \cite{Tandon:JSAC:23, Pezone:ICASSP:24, 1996SPIE:Anderson, Shlezinger:TSP:19, Zhang:IoTM:22}) since, in most AI applications, the goal is not to recover the underlying signal, but to carry out some intelligence on it. However, when the underlying communication channel is noisy, and we want to communicate under extreme bandwidth and/or power limitations, semantic and goal-oriented communications will significantly benefit from JSCC approaches to push the limits of communications for relevant applications \cite{Eirina:TCCN:19, Xie:TSP:21, Jankowski:JSAC:21, Gunduz:JSAC:23}. Moreover, many of the aforementioned applications, e.g., haptic communications and metaverse, impose strict delay constraints on communications \cite{Essaili:ETR:22}. While 1 ms. round trip delay target over the communication network has been promoted widely for 5G networks to enable such applications \cite{Fettweis:VTM:14, Simsek:JSAC:16}, these promises have not realized although ultra-low latency communication has been delivered by 5G implementations. This is due to the fact that coding latency to implement state-of-the-art compression algorithms for high-data-rate contents such as video or holograms are significantly higher than 1 ms., appearing as a major bottleneck. An important promise of JSCC approach is to reduce the coding latency by combining compression and channel coding into a single operation. Particularly, recent JSCC designs based on neural network architectures report promising results in terms of the end-to-end coding latency \cite{Eirina:TCCN:19, Kurka:IZS2020}. As we will show in the first part of this paper through information theoretic formulation, the gains from JSCC are fundamental, that is, they do not rely on a particular suboptimal implementation of separate coding approaches. In the second part, we will show practical code designs that can provide these theoretical gains in practical scenarios with an emphasis on the recent developments using deep learning techniques. }


\subsection{Outline and Organisation}

This paper provides a comprehensive overview of JSCC in communication systems, starting from information theoretic foundations to practical designs, with particular emphasis on recent progress on deep learning aided JSCC schemes.  Section \ref{s:jscc_it} is dedicated to the information theoretic foundations. \highlights{This part not only fills an important gap in the literature by providing a comprehensive overview of fundamental information theoretic performance limits and coding ideas on JSCC, but also intends to provide a foundation for the development of new neural network based codes and protocols that are inspired by these foundations.} We start with the introduction of Shannon's Separation Theorem in \ref{ss:fundamentals} for a point-to-point channel. Section \ref{ss:JSSC_Feedback} shows that the availability of channel output feedback, while does not have an impact on the optimality of separation, can allow a much simpler and low-delay JSCC scheme. We then introduce a generalized JSCC scheme when there is correlated side information at the receiver. This generalized scheme can specialize to separation-based approach with explicit source binning, as well as a simpler encoding scheme with joint decoding. We show that this joint encoding scheme can be particularly beneficial when the quality of the channel and the side information can be both uncertain. Section \ref{ss:multi-user_JSCC} is dedicated to multi-user channels. We start by showing that the separation theorem does not hold in multi-user scenarios. Then we consider multiple access and broadcast channels separately, and propose various JSCC schemes. In Section \ref{ss:SCS}, we discuss two different types of separation. In Section \ref{ss:JSCC4CC}, we show that JSCC can also help increase the rate of communication in certain multi-user scenarios. Considerations up until Section \ref{ss:beyond_average} follow Shannon's formulation: distortion between two sequences is the average distortion between pairs of symbols. In Section \ref{ss:beyond_average}, we consider distortion measures that do not necessarily satisfy this property. 

Sections \ref{s:codes_classical} and \ref{s:codes_ml} are dedicated to practical designs for JSCC. The former focuses on classical approaches, while the latter covers recent developments that rely on data-driven deep learning tools to design so-called DeepJSCC solutions. Section \ref{s:codes_ml} is organised into various subsections, each dedicated to a different source type. In particular, Section \ref{ss:image} focuses on DeepJSCC for image transmission, Section \ref{ss:video} to video transmission, Section \ref{ss:text} to text transmission, while Section \ref{ss:others} covers other source and channel distributions. Finally, Section \ref{s:conclusion} concludes the paper.




\section{Information Theoretic Foundations of JSCC}\label{s:jscc_it}


In this section, we will start with an overview of fundamental information theoretic results on JSCC, starting from Shannon's Separation Theorem, and covering various scenarios in which this theorem fails. In such cases, we also provide alternative coding schemes that can go beyond the performance achieved by pure separation, which also provide guidelines for the design of practical codes.

\subsection{Fundamentals}\label{ss:fundamentals}

\begin{figure}
		\centering
		\includegraphics[width=3.5in]{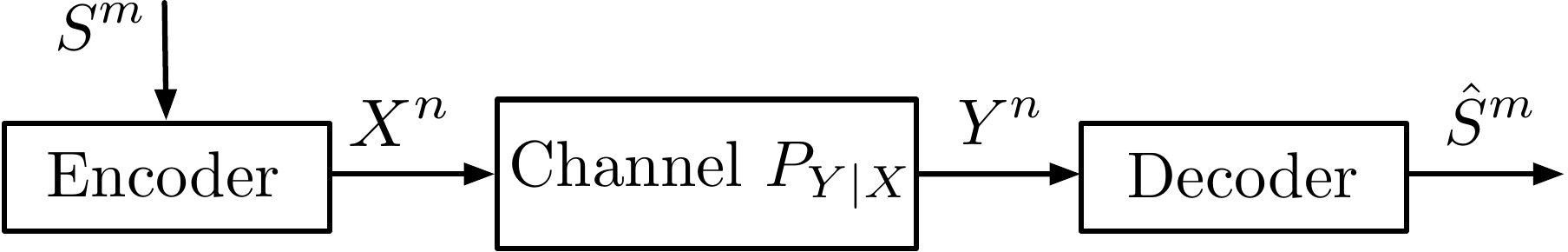}
		\caption{Illustration of a JSCC problem over a noisy communication channel.}
  \label{f:JSCC_model}
	\end{figure}

In the fundamental problem of information theory as studied by Shannon in \cite{Shannon}, the transmitter wants to transmit the output of an information source to a receiver over a noisy communication channel. In general, the goal of the receiver is to recover the observed source sequence either reliably (i.e., with vanishing error probability), or within some distortion limit under a prescribed distortion measure. Shannon formulated this problem using a statistical model for the source signal and the channel transform, and focused on a block coding framework in the asymptotic infinite blocklength regime. In particular, the model considered by Shannon is illustrated in Fig.~\ref{f:JSCC_model}, and described next in a mathematical form. 

Assume that the encoder wants to transmit a sequence of independent source symbols $S^m \in \mathcal{S}^m$ sampled from the source distribution $p_S(s)$, over a memoryless noisy communication channel characterized by the conditional probability distribution $P(Y|X)$, where $X \in \mathcal{X}$, $Y \in \mathcal{Y}$. Let $\hat{S}^m \in \mathcal{\hat{S}}^m$ denote the reconstruction at the receiver based on $Y^n$. Similarly to the rate-distortion theory formulation, the goal is to minimize the distortion between $S^m$ and $\hat{S}^m$ under some given distortion (fidelity) measure, $d: \mathcal{S}^m \times \mathcal{\hat{S}}^m \rightarrow [0, \infty)$. More formally, let $f^{m,n}:  \mathcal{S}^m \rightarrow \mathcal{X}^n$ denote the encoding function, and $g^{m,n}:  \mathcal{Y}^n \rightarrow \mathcal{\hat{S}}^m$ denote the decoding function. In the case of an average distortion criteria, the goal is to identify the encoder and decoder function pairs that minimize $\mathds{E}[d(S^m, \hat{S}^m)]$, where the expectation is over the source and channel distributions as well as any randomness the encoding and decoding functions may introduce.  One can also impose an excess distortion criterion, where the goal is to minimize $\mathds{P}[d(S^m, \hat{S}^m) >D]$, for some maximum allowable distortion target $D>0$. 

An $(m,n)$ joint source-channel code of rate $r=m/n$ consists of an encoder-decoder pair, where the encoder $f^{(m,n)}: \mathcal{S}^m \rightarrow \mathcal{X}^n$ maps each source sequence $s^m$ to a channel input sequence $x^n(s^m)$, and the decoder $g^{(m,n)}: \mathcal{Y}^n \rightarrow \mathcal{\hat{S}}^m$ maps the channel output $y^n$ to an estimated source sequence $\hat{s}^m$. 

\begin{definition}\label{d:achieve}
A rate-distortion pair $(r,D)$ is said to be \textit{achievable} if there exists a sequence of $(m,n(m))$ joint source-channel codes with rate $r$ such that $r \cdot n(m) \leq m$, $\forall m$, and 
\begin{equation}
	\limsup_{m\rightarrow \infty} \mathrm{E}[d(S^m, \hat{S}^m(Y^{n(m)}))] \leq D.
\end{equation}
\end{definition}

Similarly to the definition of channel capacity, we can define a source-channel capacity for a given distortion target, which considers both the source and the channel characteristics.

\begin{definition}\label{d:capacity}
For a given target distortion $D$, the \emph{source-channel capacity} of a channel is defined as the supremum of rates $r$ among all achievable $(r,D)$ pairs.
\end{definition}

Shannon proved his Separation Theorem considering single-letter additive distortion measures; that is, 
\begin{equation}\label{eq:add_Dist}
d(S^m, \hat{S}^m) = \frac{1}{m} \sum_{i=1}^m d(S_i, \hat{S}_i)
\end{equation}
for the distortion measure $d(S, \hat{S}) \in [0, \infty)$. The Separation Theorem can be stated as follows, \highlight{after defining the capacity of a discrete memoryless channel $P_{Y|X}$:
\begin{equation}
C :=  \sup_{p_X(x)} I(X;Y),
\end{equation}
and the rate-distortion function of a memoryless source $p_S$:
\begin{equation}
R(D) := \inf_{ P_{\hat{S}|S}} I(S;\hat{S}),
\end{equation}
where the infimum is taken only over conditional distributions satisfying $\mathrm{E}[ d(S,\hat{S})] \leq D$.}

\begin{theorem}\label{thm1}(Shannon's Separation Theorem, \cite{Shannon}) Given a memoryless source $S$ with distribution $p_S$ and a memoryless channel $p_{Y|X}$ with capacity $C $, a rate-distortion pair $(r,D)$ is achievable if $r R(D) < C$. Conversely, if a rate-distortion pair $(r,D)$ is achievable, then $r R(D) \leq C$. Then, for given distortion target $D$, the source-channel capacity is given by $C/R(D)$. 
\end{theorem}
	\begin{figure}
		\centering
		\includegraphics[width=3.5in]{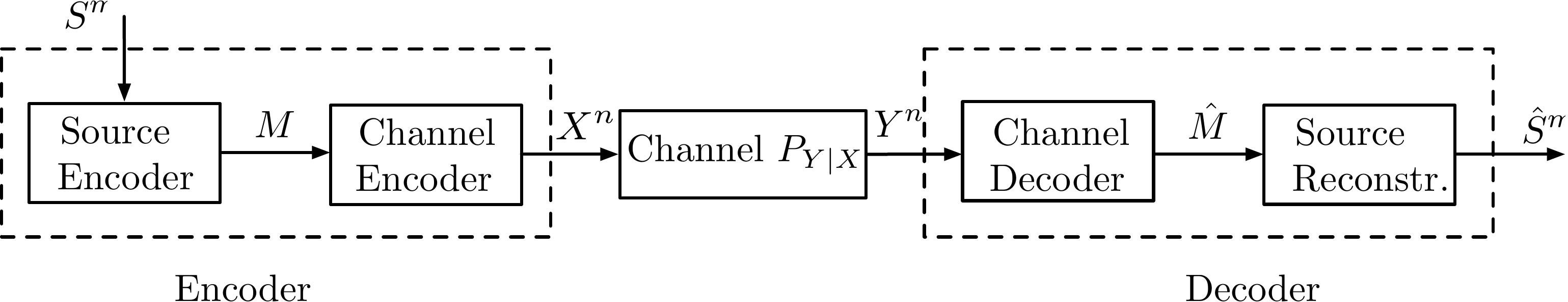}
		\caption{Separation-based architecture for the JSSC problem.}
  		\label{fig:separation}
	\end{figure}

The theorem states that we can separate the design of the communication system into two sub-problems without loss of optimality, the first focusing on compression and the second focusing on channel coding, each of them designed independently of the other. This separation-based architecture is depicted in Fig.~\ref{fig:separation}. However, the optimality of separation holds only in the limit of infinite blocklength \highlight{and only in terms of capacity. For example, still in the asymptotic regime of infinite blocklengths, the exponential decay-rate of the probability of error  as well as the second-order coding rates can be improved with a joint source-channel code compared to the best separation-based approach, see \cite{Gallager:book,Csiszar:PCIT:1980,Zhong:TIT:2006, Kostina:TITL2013}. The same observation also holds in the  regime of finite blocklengths, where optimal joint source-channel codes outperform the best  separation-based code design.}

The suboptimality of separation in the general finite blocklength regime was observed by Shannon in his 1959 paper \cite{Shannon_RD}, where he considered a binary source generating independent and equiprobable symbols and a memoryless binary symmetric channel, and  observed that simple uncoded transmission of symbols achieves the optimal distortion with rate $r=1$ for one particular value of distortion $D$ determined by the error probability of the channel. This observation was later extended by Goblick in \cite{Goblick:TIT:65} to Gaussian sources transmitted over Gaussian channels. 

In general, \highlight{ uncoded transmission is optimal   \cite{Gastpar:TIT:03}} when the source distribution matches the optimal capacity achieving input distribution of the channel, and the channel at hand matches the optimal test channel achieving the optimal rate-distortion function of the source. \highlight{I.e, when  the  source and channel input alphabets match, $\mathcal{X}=\mathcal{S}$ and the source distribution $p_S$ and the channel transition law $P_{Y|X}$ satisfy the following two conditions:
\begin{eqnarray}
    I(X;Y) |_{X\sim p_S} &= & C\\
    I(S;\hat{S})|_{\hat{S}\sim P_{Y|X}(\cdot|S)} &= & R(D).
\end{eqnarray}} However, these conditions are not satisfied for most practical source and channel distributions, and even when they hold, optimality of uncoded transmission fails when the coding rate is not $1$, i.e., in the case of bandwidth compression or expansion. On the other hand, the presence of such optimality results, that is, the fact that asymptotically optimal performance, which requires infinite blocklength source and channel codes in general, can be achieved by simple zero-delay uncoded transmission implies that there can be other simple joint coding schemes that can achieve near optimal performance, and outperform separation-based schemes in the finite blocklength regime. Various   JSCC schemes that perform well in the finite blocklength regime have been proposed, see for example,  in \cite{Kostina:TITL2013,Kochman:ISIT:2019, Truong:TIT:2019,Zheng:ISTC:2021}.

Interesting phenomena have also been proved for JSCC setups that are not memoryless, in particular also for the special case where the source has to be reconstructed perfectly. While the separation-theorem has been shown to continue to hold for a large class of sources and channels \cite{Dob59, Hu:ITCz:1962, Aghajan:Prob:2015}, the authors in  \cite{Vembu:TIT:1995, Han:book:2014} have identified source-channel pairs where separation does not hold.  In fact, they  characterized matching necessary and sufficient conditions for a source to be transmittable over a channel for a large class of source distributions and channel law. Uncoded transmission, and thus JSCC coding, can also be advantageous under channel uncertainty, for example over fading channels, where in certain cases (e.g., for the transmission of Gaussian sources over a Gaussian fading channel with receiver channel state-information) it may allow attaining the optimal distortion under any given channel condition. SSCC scheme can either be in excess, and thus, result in huge distortion under bad channel conditions, or has to rely on more advanced and complex coding schemes (like the multi-layer broadcast approach of \cite{Shamai:IT:03, Gunduz:IT:07, Gunduz:TIT:08a}).

\subsection{JSCC with a Remote Source}\label{ss:JSSC_Remote}

\begin{figure}
		\centering
		\includegraphics[width=3.5in]{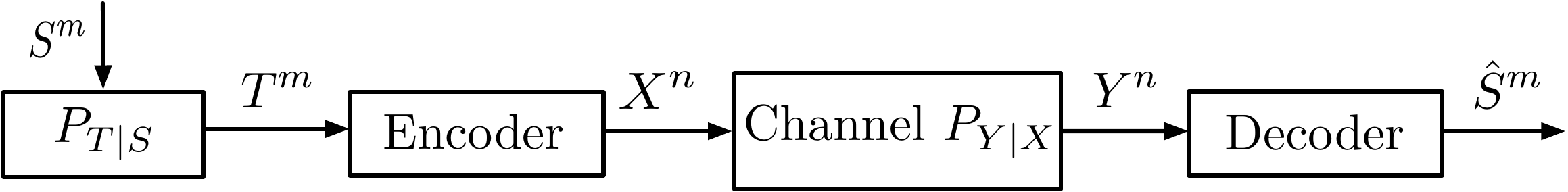}
				\caption{Illustration of a JSCC problem with a remote source $S^m$ that has to be reconstructed at the decoder.}
  \label{fig:JSCC_remote}
	\end{figure}

\highlight{
A variation of the standard source coding problem that is particularly relevant to semantic communication is the setup of remote JSSC depicted in Fig.~\ref{fig:JSCC_remote}. In this setup,  the encoder does not directly have access to the source sequence itself but only to a related observation $T^m$. The remote source could for example represent a sequence of features of the image $T^m$. The main leitmotif of   semantic communication is that the decoder does not wish to reconstruct the entire image but only the sequence of features. In the remote JSCC problem the decoder indeed attempts to reconstruct the remote (hidden) source sequence $S^m$ and not the sequence observed at the encoder. }

\highlight{
Optimality of source-channel separation for the remote JSCC has been established independently by Dobrushin and Tsybakov  \cite{Dobrushin} and by Wolf and Ziv \cite{Wolf:TIT:70}. More recently, extensions of this problem has been considered in \cite{Liu:TCOM:22} and \cite{Stavrou:TCOM:23}, in which both the remote and the observed sources are reconstructed at the receiver. For this setup, Theorem~\ref{thm1} has to be adapted by employing the \emph{remote rate-distortion function}:
\begin{equation}
R_{\textnormal{remote}}(D) = \inf_{ p_{\hat{S}|T}} I(T;S),
\end{equation}
where the infimum is taken over all laws $p_{\hat{S}|T}$ satisfying the distortion constraint $\mathrm{E}[ d(S, \hat{S})] \leq D$.}
\begin{theorem}[Separation Theorem with a Remote Source] Given a remote memoryless source $S$ with distribution $p_S$, a memoryless observation $T$ related to the source by the transition law $p_{T|S}$,   and a memoryless channel $p_{Y|X}$ with capacity $C $, a rate-distortion pair $(r,D)$ is achievable if $r R_{\textnormal{remote}}(D) < C$. Conversely, if a rate-distortion pair $(r,D)$ is achievable, then $r R_{\textnormal{remote}}(D) \leq C$. Thus, for given distortion target $D$, the source-channel capacity is given by $C/R_{\textnormal{remote}}(D)$. 
\end{theorem}
 \highlight{
The optimal SSCC scheme can be described as follows. The encoder attempts to reconstruct the hidden features $S^m$ from the observed sequence $T^m$ with desired accuracy, and then uses a capacity-achieving channel code  to send this information reliably (i.e., with vanishing probability of error) to the decoder, which first decodes the transmitted information and then reproduces the encoder's reconstruction $\hat{S}^m$.
}
 
\subsection{JSCC over Feedback Channels}\label{ss:JSSC_Feedback}

Optimality of source-channel separation (Theorem~\ref{thm1}) extends also to setups with feedback, i.e., to setups where the transmitter  observes the past channel outputs $Y_1, \ldots, Y_i$ (or noisy or imperfect versions thereof)  before producing the time-$i$ channel input $X_i$. Thus, also with feedback,  a rate-distortion pair $(r,D)$ is achievable if, and only if, $r R(D) < C$. Notice here that, for memoryless point-to-point channels feedback does not increase the capacity; therefore, the best achievable source-channel coding rate remains the same with feedback.

JSCC, however, allows to reduce the delay and obtain simplified schemes. We illustrate this on the well-known example of sending a Gaussian source over a Gaussian channel with a bandwidth mismatch factor $r$, whose inverse $r^{-1}$ is a positive integer. The proposed scheme is based on an idea of Schalkwijk and Kailath \cite{Schalkwijk:TIT:66}.  A similar scheme can be designed for arbitrary discrete sources and discrete memoryless channels using the `posterior matching' idea \cite{Shayevitz:TIT:11}. 

Let  $S^m$ be a sequence of independent and identically distributed (i.i.d.) centered Gaussian samples of variance $\sigma^2$. Consider further an additive white Gaussian noise (AWGN) channel with input-output relation
\begin{equation}
    Y_i = X_i + Z_i,\quad i=1,\ldots, n, \label{AWGN_channel}
\end{equation}
where $\{Z_i\}$ is an i.i.d. sequence of centered Gaussian noise variables of variance $N$. We assume perfect channel output feedback, in which case $X_i$ can depend on the past channel outputs $Y^{i-1}$. Channel inputs are subject to an expected average block-power constraint $\frac{1}{n} \sum_{i=1}^n \mathbb{E}[X_i^2 ]\leq P$, for a given $P>0$.  For this Gaussian-quadratic setup distortion $D$ is achievable if, and only if\footnote{Since the right-hand side of \eqref{eq:condition} is non-negative, the left-hand side can be simplified to $\frac{1}{2}\log\frac{\sigma^2}{D}$.} 
\begin{equation}\label{eq:condition}
 \max\left\{ \frac{1}{2} \log \left( \frac{ \sigma^2}{D} \right), 0 \right\}\leq \frac{1}{2} \log\left( 1+ \frac{P}{N}\right),
\end{equation}
where notice that the left-hand side of above inequality is the Gaussian-quadratic rate-distortion function and the right-hand side  the capacity of the power-constrained Gaussian channel.

The following low-delay scheme achieves distortion $D$ whenever \eqref{eq:condition} is satisfied. The encoder communicates each source  $S_t$ over $r^{-1}$ consecutive channel uses. We describe the transmission of $S_1$ during the first $r^{-1}$ channel uses. The other transmissions are similar. 

The encoder first sends 
\begin{equation}
    X_1 =\sqrt{\frac{P}{\sigma^2}} S_1
\end{equation}
and subsequently observes the output symbol $Y_1=X_1+Z_1=\sqrt{\frac{P}{\sigma^2}} S_1+Z_1$. From this output it computes the minimum mean-squared error (MMSE) estimate  of source symbol $S_1$:
\begin{equation}
\tilde{S}_{1,1}= \frac{\sqrt{P\sigma^2}}{P+N} Y_1 =\frac{P}{P+N} S_1+  \frac{\sqrt{P\sigma^2}}{P+N} Z_1
\end{equation}   and the corresponding MMSE 
\begin{equation}
    \epsilon_1 = \tilde{S}_{1,1}-S_1.
\end{equation}
It then scales this error so as to meet the input power constraint and transmits this error over the next channel input: 
\begin{equation}
    X_2= \sqrt{\frac{P}{\alpha_1}} \epsilon_1, 
\end{equation}
for $\alpha_1:= \mathbb{E}[\epsilon_1^2]$.
After reception of the feedback output $Y_2$, the encoder computes the  MMSE estimate of $\epsilon_1$: 
\begin{equation}
\tilde{\epsilon}_1= \frac{\sqrt{P\alpha_1 }}{P+N} Y_2 =\frac{P}{P+N} S_2+  \frac{\sqrt{P\alpha_1}}{P+N} Z_2
\end{equation} and uses it to update its estimate on $S_1$: 
\begin{equation}
    \tilde{S}_{1,2} = \tilde{S}_{1,1}- \tilde{\epsilon}_1.
\end{equation}
The new estimation error then becomes 
\begin{equation}
    \epsilon_2 =    \tilde{S}_{1,2}  - S_1= \tilde{\epsilon}_1-\epsilon_1, 
\end{equation}
and is sent over the channel with the next input $X_3$.
These steps are repeated until $r^{-1}$ channel inputs are sent. Specifically, for any $\ell=2,3,\ldots, r^{-1}$, after sending the latest estimation error 
\begin{equation}
    X_\ell=\sqrt{\frac{P}{\alpha_{\ell-1}}}\epsilon_{\ell-1}, \quad \ell= 2,3,\ldots, r^{-1},
    \end{equation}
    at time $\ell$, where 
    \begin{equation}
    \alpha_{\ell-1}:=\mathbb{E}[\epsilon_{\ell-1}^2],
    \end{equation}
    the encoder uses the feedback $Y_\ell=X_\ell+Z_\ell$ to calculate an estimate 
    \begin{equation}
    \tilde{\epsilon}_{\ell-1}= \frac{\sqrt{P\alpha_{\ell-1} }}{P+N} Y_{\ell-1} ,
    \end{equation} which is then used to update the estimate $\tilde{S}_{1,\ell}$ of $S_1$: 
\begin{equation}
\tilde{S}_{1,\ell}=\tilde{S}_{1,\ell-1}-\tilde{\epsilon}_{\ell-1} = S_1 + \epsilon_{\ell-1}- \tilde{\epsilon}_{\ell-1}
\end{equation}
Subsequently, a scaled version of the estimation error 
\begin{equation}\epsilon_\ell:= \tilde{S}_{1,\ell}-S_1= \epsilon_{\ell-1}-\tilde{\epsilon}_{\ell-1}
\end{equation}
is sent in the following transmission $X_{\ell+1}$.

After receiving the first $r^{-1}$ channel outputs $Y_1,\ldots, Y_{r^{-1}}$, the receiver sets 
\begin{equation}\label{eq:estimate}
    \hat{S}_1=\tilde{S}_{1,r^{-1}},
\end{equation}
where it computes $\tilde{S}_{1,r^{-1}}$ as described for the encoder-side.

\begin{figure}
		\centering
		\includegraphics[width=3.5in]{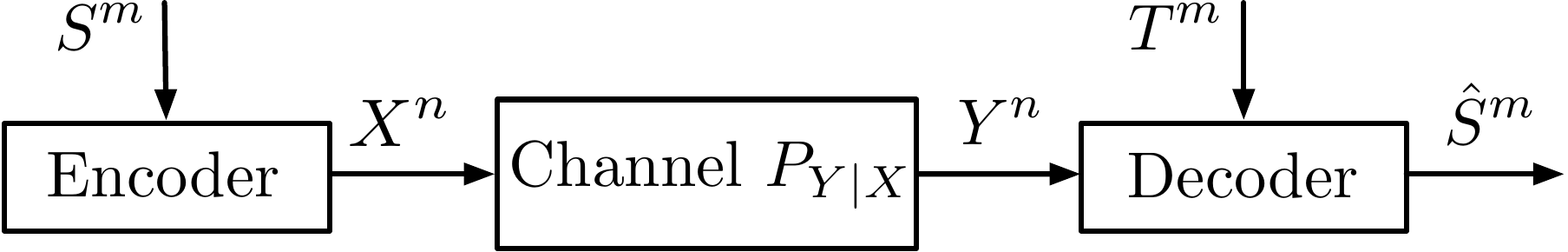}
		\caption{JSCC problem in the presence of correlated side information at the receiver.}
  		\label{fig:SW_JSCC}
\end{figure}

To derive the distortion achieved by \eqref{eq:estimate}, notice that the MMSE between $\tilde{S}_{1,\ell}$ and $S_1$ can recursively be  calculated as 
\begin{eqnarray}
   \mathbbm{E} \left[  (\tilde{S}_{1,\ell}-S_1)^2 \right] &= & \alpha_{\ell} = 
 \frac{N}{P+N}\alpha_{\ell-1},
\end{eqnarray}
and thus $\alpha_\ell = \sigma^2 \left( \frac{N}{P+N}\right)^\ell$. The proposed coding scheme thus matches MMSE distortion constraint $d$ whenever 
\begin{equation}
  \left( \frac{N}{P+N}\right)^\ell \leq \frac{D}{\sigma^2}, 
\end{equation}
which is equivalent to \eqref{eq:condition} when specialized to $\ell=r^{-1}$ because $\frac{N}{P+N}<1$ and because of the monotonicity of the log-function. 

Compared to a SSCC scheme, the above scheme is much simpler and has a reduced delay, that is, each sample can be decoded after $1/r$ channel uses, instead of waiting the transmission of all the channel symbols to decode all the source symbols at once.

\subsection{JSCC with Side Information: To Bin or Not To Bin?} \label{ss:SW_JSCC}

Next, we consider the scenario in which the receiver has access to correlated side information (see Fig. \ref{fig:SW_JSCC}), that is, samples of $(S_i, T_i)$ come from a joint distribution $p(s,t)$. Here we will focus on lossless transmission of $S^m$ to highlight the core idea of binning; the extension to lossy transmission follows similarly. Different from the scenario in Section \ref{ss:fundamentals}, here the decoder has access to correlated side information $T^m$, so the decoding function is now given by $g^{(m,n)}: \mathcal{Y}^n \times \mathcal{T}^m \rightarrow \mathcal{S}^m$. From a source coding perspective, this is known as the Slepian-Wolf coding problem \cite{Slepian:BSTJ:73}. 

This problem was originally studied by Shamai and Verd\'{u} in \cite{Shamai:ETT:95}, where it is shown that a source-channel code rate $r$ is achievable if there exists an input distribution $p(x)$ for which $r < \frac{I(X;Y)}{H(S|T)}$; and conversely, if $r$ is achievable then there exists an input distribution $p(x)$ for which $r \leq \frac{I(X;Y)}{H(S|T)}$. Equivalently, the source-channel capacity of this system is given by $\frac{C}{H(S|T)}$, where $C$ is the capacity of the underlying channel. It is not difficult to see that the Separation Theorem continues to hold in this setting; that is, any source-channel code rate less than the source-channel capacity, $\frac{C}{H(S|T)}$, can be achieved by first applying source coding, and then transmitting the compressed source bits over the channel using a capacity-achieving channel code. 

The source encoder in the case of separation employs Slepian-Wolf coding, which exploits \emph{binning} of the source outcomes. The source encoder randomly distributes all possible source output sequences $S^m$ into $2^{mH(S|T)}$ bins; that is, it independently assigns an index uniformly distributed over $[1: 2^{mH(S|T)}]$ to each possible source sequence. This constitutes the compression codebook. We also generate a channel codebook of the same size, consisting of length-$n$ channel codewords, each generated in an i.i.d. fashion from $\prod_{t=1}^n p(x_{t})$, where $p(x)$ is the capacity achieving input distribution. Then, to transmit the bin index to the decoder, corresponding channel codeword is transmitted over the channel. Since $r< \frac{C}{H(S|T)}$, the transmitted channel codeword will be decoded correctly with high probability. Having decoded the channel codeword, the decoder recovers the bin index, and outputs the source sequence in the corresponding bin that is jointly typical with its side information sequence, $T^m$. It is shown in \cite{Slepian:BSTJ:73} that the decoder can recover the correct source sequence with high probability since the number of bins is at least $2^{mH(S|T)}$.

Next, we present a coding scheme that generalizes SSCC \cite{Gunduz:TIT:13b}. In this generalized coding scheme, we randomly distribute all $S^m$ sequences into $M=2^{mR}$ bins, where $R$ is not necessarily equal to $H(S|T)$. Let $\mathcal{B}(i)$ be the set of sequences allocated to bin $i$. Then, we generate $M$ i.i.d. length-$n$ channel codewords with distribution $\prod_{t=1}^n p(x_{t})$, and enumerate these codewords as $x^n(w)$ for $w=1, \ldots, M$. This constitutes the only codebook in the system. Encoding is done similarly to the separation-based scheme. The transmitter finds the index $i$ of the bin $s^m$ belongs to, and transmits the corresponding codeword $x^n(i)$. 

Note that, in the separation approach described above, the channel decoder decides a single channel codeword index, which conveys the bin index that is then used for source decoding. In the separation approach, for the channel decoding to be successful, it is better to reduce the number of possible indices to be transmitted. In the generalized scheme, we consider a joint source-channel decoder, following the approach in \cite{Tuncel:TIT:2006}. The decoder looks for an index $i$ for which $x^n(i)$ and $Y^n$ are jointly typical; and, at the same time, there exists exactly one source sequence $\hat{s}^m$ in bin $i$ that is jointly typical with its side information, $T^m$. 


In the joint decoding scheme, we have an error if there exists no or more than one such bin index $i$, or if there exist more than one jointly typical sequences within bin $i$. The probability that there is no bin index satisfying the joint typicality condition vanishes as $n$ grows. The probability of having no jointly typical source sequence within the correct bin also vanishes since $S^m$ and $T^m$ are jointly typical with high probability as $m$ grows. Using the classical arguments on typical sets \cite{Csiszar:book}, the probability of having another jointly typical source sequence in the same bin as $S^m$ can be bounded by
\begin{equation}\label{eq:PtP:1}
\begin{aligned}
\big|\mathcal{B}(i) \bigcap A_\epsilon^m(S) \big| & 2^{-m(I(S; T)-3\epsilon)}  \\ 
&~~~~~~~ \leq 2^{m(H(S)+\epsilon)} 2^{-mR} 2^{-m(I(S;T)-3\epsilon)},  
\end{aligned}
\end{equation}
where $A_\epsilon^m(S)$ denotes the set of $\epsilon$-typical $m$-tuples according to $P_{S}$. We can see that (\ref{eq:PtP:1}) goes to zero if $R \geq H(S|T)$.

We also have an error if there exists another bin index $j$ satisfying the joint typicality conditions. The probability of this event can be bounded by
\begin{equation}
\begin{aligned}
    2^{mR} 2^{-n(I(X; Y)-3\epsilon)} |\mathcal{B}(i) \bigcap A_\epsilon^m(S)|2^{-m(I(S; T)-3\epsilon)} \\ 
    ~~~~ \leq 2^{-n(I(X; Y_1)-3\epsilon)} 2^{m(H(S|T)-2\epsilon)},
\end{aligned}
\end{equation}
which goes to zero if $mH(S|T) < nI(X ; Y_1)$. Hence, any rate $r$ satisfying $r <\frac{I(X; Y)}{H(S|T)}$ is achievable.

We have obtained a set of coding schemes each with a different number of bins, that is, with different $R$ values satisfying $R \geq H(S|T)$. The ``joint'' decoding operation considered in the generalized scheme can equivalently be viewed as a separate source and channel decoding scheme, in which the channel decoder is a list decoder, which outputs the list of bin indices $i$ for which $x^n(i)$ and $Y^n$ are jointly typical. This list decoding approach includes SSCC as a special case with $R=H(S|T)$, in which case, with high probability, there is a single element in the list, i.e., there exists only a single bin index whose channel codeword is typical with the channel output.

Please also note that, in the other extreme, this generalized scheme works without any binning at all. We can generate an independent channel codeword for each possible source outcome, i.e., $R = \log |\mathcal{S}|$. From a practical point of view, this can be interpreted as transferring the complexity of binning from the encoder to the decoder, which now needs to apply joint decoding or list decoding instead of separate source and channel decoding steps. From a theoretical point of view, since the decoder only outputs typical source sequences, there is no point in having more than $2^{m(H(S)+\epsilon)}$ bins as, otherwise, we are creating bins without any typical source sequence in them; and hence, they will never be output. Therefore, we can set $R \leq H(S)$ without loss of generality. 

In the case of a point-to-point channel, the only difference between SSCC with binning and joint decoding with no-binning is the operation at the encoder and the decoder. However, binning or no-binning can lead to different protocols, or even different performances in the case of multi-user networks or non-ergodic settings as we will see in the next section. Even in the point-to-point scenario considered here, if the channel and the side information are time-varying, and if the transmitter knows only the distributions of these time-varying processes, it is shown in \cite{Aguerri:TIT:16} that the no-binning scheme improves the end-to-end performance compared to separate binning and channel coding, as the former naturally adapts to the source and channel states without targeting specific source and channel realizations.


\subsection{Multi-user JSCC}\label{ss:multi-user_JSCC}

\subsubsection{Suboptimality of SSCC with Correlated Sources}

\begin{figure}
		\centering
		\includegraphics[width=3.5in]{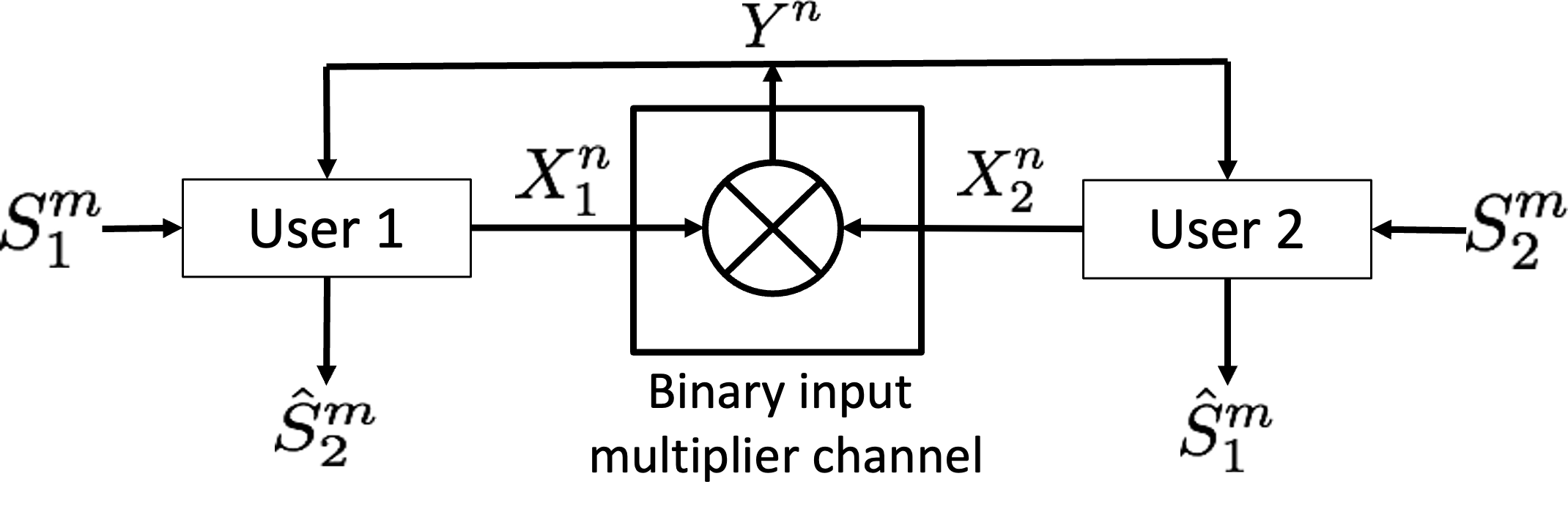}
		\caption{JSCC problem over the binary multiplier two-way communication channel.}
  		\label{fig:2way_JSCC}
\end{figure}

Shannon's Separation \highlight{paradigm (Theorem~\ref{thm1}) extends to a certain class of problems where multiple sources are transmitted over a  network with multiple transmitters and/or multiple receivers. In particular, source-channel separation is optimal for: 
\begin{itemize}
    \item 
all problems where all source sequences are   memoryless and  independent of each other and each source sequence has to be reconstructed at a single receiver;  and 
\item  all problems where the network can be modeled as a set of non-interfering (orthogonal) point-to-point links.
\end{itemize} The paradigm however does not extend to general multi-terminal networks with arbitrarily correlated sources. In general, for these networks  JSCC achieves a better performance.} A particular advantage of JSCC in the multi-user context is that it allows to transfer the correlation between the sources observed at different terminals  to their channel inputs, which is not possible via SSCC, since the latter relies on transmitting independently generated channel codewords. 

In the literature, the proof of the suboptimality of separation is commonly attributed to the work by Cover, El Gamal and Salehi \cite{Cover:TIT:80}, which studies the transmission of correlated sources over a multiple access channel (MAC). However, Shannon already made the observation that JSCC can outperform separation in \cite{Shannon1961TwowayCC} in the context of two-way communication. In this paper, Shannon proposed inner and outer bounds on the set of achievable rate pairs over a general two-way communication channel, whose capacity remains an open problem to this day. Shannon particularly considered the multiplier channel, previously introduced by David Blackwell, as an example, which is illustrated in Fig. \ref{fig:2way_JSCC}. In this channel, both transmitters have binary inputs, i.e., $\mathcal{X}_1= \mathcal{X}_2 = \{0,1\}$, while the common channel output is also binary and given by $y = x_1 x_2$. Shannon first showed numerically that his inner and outer bounds do not match for the multiplier channel. Then, by considering correlated source sequences $S_1^m$ and $S_2^m$ available at the two encoders, where $p_{S_1S_2}(s_1, s_2)$ satisfies $p(0,0) = 0$, $p(0,1) = p(1,0) = 0.275$ and $p(1,1) = 0.45$, Shannon showed that it is possible to reach the outer bound on the equal rate point. Note that, with this particular source distribution no coding is needed, and by simply sending the source realizations over the channel, each user can recover the other user's source symbol reliably, and hence, a source-channel code rate of $r=1$ source sample per channel use is achievable. Shannon did not explicitly show that SSCC cannot achieve this source-channel rate performance. Transmitting these sources at the same source-channel code rate with SSCC would require the two users simultaneously transmit at rates $H(S_1|S_2) = H(S_2|S_1) = 0.6942$ bits per channel use. A tighter outer bound on the capacity region of the two-way channel was proposed by Hekstra and Willems in \cite{Hekstra:TIT:89}. According to the dependence-balance bound derived in \cite{Hekstra:TIT:89}, the symmetric rate values larger than $R_1 = R_2 = 0.64628$ bits per channel use are not achievable. Therefore, we can conclude that the correlated sources $S_1$ and $S_2$ specified above, which can be easily transmitted over the multiplier two-way channel without any coding, cannot be reliably transmitted by any channel code. This result shows that the Separation Theorem does not generalize to arbitrary multi-user channels. 

\begin{figure}
		\centering
		\includegraphics[width=3.5in]{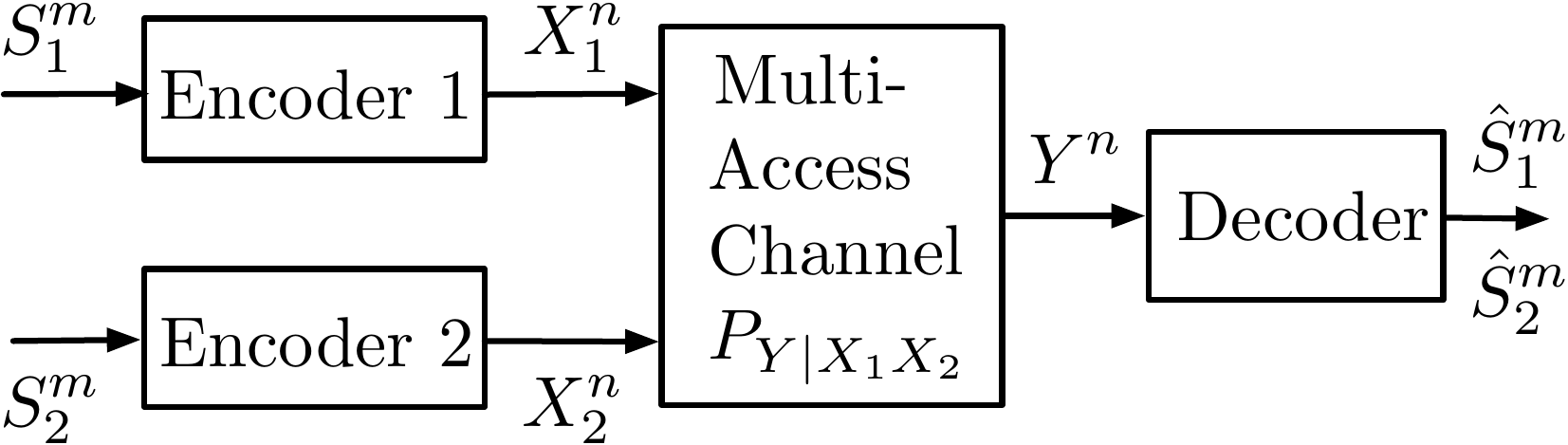}		\caption{Transmission of correlated sources over a multi-access channel (MAC).}
  		\label{fig:correlated_MAC}
\end{figure}

\subsubsection{Correlated Sources over a Multiple-Access Channel (MAC)}

As we have mentioned above, a better known example illustrating the suboptimality of separation in multi-user scenarios considers the transmission of correlated sources over a MAC, see  Fig.~\ref{fig:correlated_MAC}. Cover, El Gamal and Salehi showed the suboptimality of separation through an example very similar to the one considered by Shannon for the two-way channel. They considered a binary-input adder channel, where the channel output is given by $Y = X_1 + X_2$, $\mathcal{X}_1= \mathcal{X}_2 = \{0,1\}$ and $\mathcal{Y} = \{0, 1, 2\}$, and a source distribution $p_{S_1S_2}(s_1, s_2)$ characterized by $p(0,0) = 0$, $p(0,1) = p(1,0) = p(1,1) = 1/3$. Similarly to Shannon's example, uncoded transmission of the source samples trivially achieves a source-channel rate of $r=1$. On the other hand, unlike the two-way channel example of Shannon, the capacity region of this MAC is known, and the maximum sum rate that can be achieved by independent channel inputs is $1.5$ bits per channel use, while the joint entropy of the two sources is $H(S_1, S_2)= \log 3 = 1.58$ bits per sample. Hence, the maximum source-channel rate that can be achieved through separation is given by $1.5/1.58 \sim 0.949$ channel uses per sample.

\begin{figure*}
		\centering
		\includegraphics[width=5in]{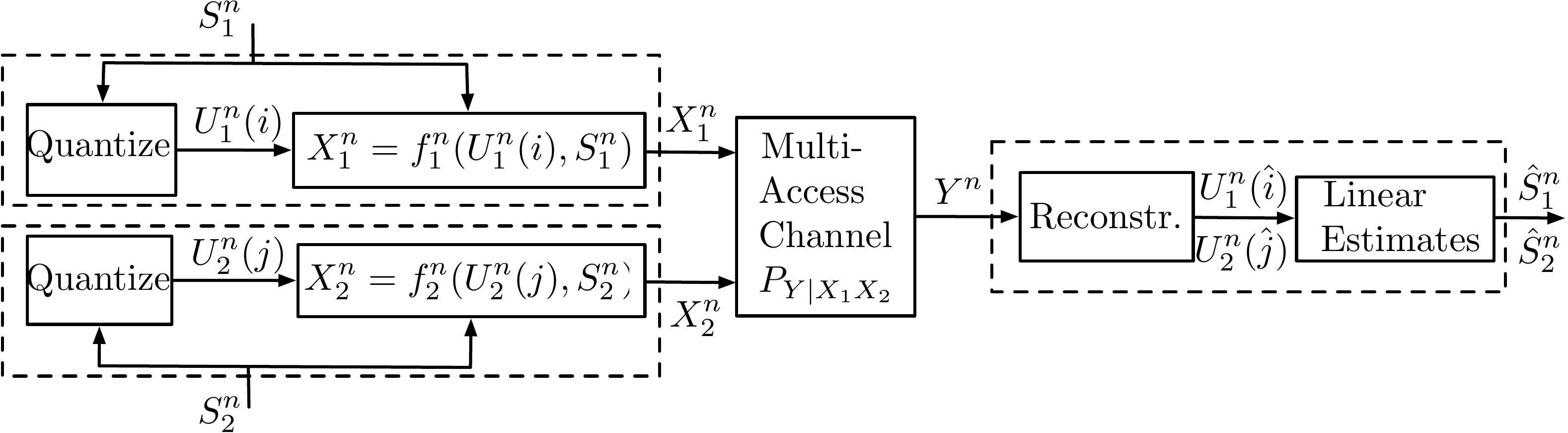}
		\caption{Hybrid coding for transmitting a bivariate source over a multi-access channel. }
  		\label{fig:hybrid_MAC}
\end{figure*}

JSCC strategies and achievability results for general discrete memoryless sources and MACs were reported  for the lossless case \cite{Cover:TIT:80} as well as for the lossy case \cite{Salehi:ISIT:95, Minero:IT:15}. Information theoretic converses were established in  \cite{Kang:CISS:06, Lapidoth:ISIT:2017, Guler:IT:18}. In the finite blocklength regime, i.e., for fixed and non-asymptotic $n$, JSCC schemes where proposed in \cite{Padakandla2:IT:21}.

For Gaussian sources with squared-error distortions and Gaussian MACs without bandwidth expansion ($r=1$), \cite{Tinguely:TIT:10} presented both  JSCC strategies and corresponding achievability results, as well as a (partially matching and generally very tight) converse result. We describe the results for the Gaussian setup in \cite{Tinguely:TIT:10} in more detail.  Consider the communication setup in Fig. \ref{fig:correlated_MAC}, where two users observe correlated jointly Gaussian source sequences $S_1^n$ and $S_2^n$, which they wish to communicate to a common receiver over $n$ uses of a Gaussian MAC with input-output relation $Y_t=X_{1,t}+X_{2,t}+Z_t$. Here $Z_t$ are independent samples from a zero-mean Gaussian distribution with variance $N$, and the average block-power constraints of $P_1$ and $P_2$ are imposed on the channel inputs of the users. We thus have $m=n$ and no bandwidth mismatch in this scenario, i.e., $r=1$. Based on the observed sequence of channel outputs $Y^n$, the receiver has to produce reconstruction sequences $\hat{S}_1^n$ and $\hat{S}_2^n$ that satisfy the squared-error distortion constraints $\mathbb{E}[ \|\hat{S}_k^n- S_k^n\|^2 ] \leq D_k$, for $k\in\{1,2\}$.

Consider a symmetric setting where the symbols $(S_{1,t}, S_{2,t})$ are assumed i.i.d. jointly  Gaussian of variances $\sigma^2$ and correlation coefficient $\rho$, and we set the power constraints equal, i.e., $P_1=P_2=P$. It is shown in \cite{Tinguely:TIT:10} that, for large  correlation factors $\rho$ satisfying $\frac{\rho}{1-\rho} \geq \frac{P}{N}$, uncoded transmission achieves the minimum possible distortion values $D_1$ and $D_2$, i.e., the best encodings are to choose $X_{1,t}= \sqrt{\frac{P}{\sigma^2}} S_{1,t}$ and $X_{2,t}= \sqrt{\frac{P}{\sigma^2}} S_{2,t}$ and optimal  decoding is to produce the linear MMSE of $S_{1,t}$ and $S_{2,t}$ based on channel output $Y_t$, for each $t=1, \ldots, n$. Separation-based schemes can be shown to achieve significantly larger distortions, and thus worse performance.

For  correlation coefficients $\rho$ satisfying $\frac{\rho}{1-\rho} \geq \frac{P}{N}$, the hybrid coding scheme in Fig. \ref{fig:hybrid_MAC} can achieve smaller distortions than uncoded transmission. The idea of hybrid coding is that each encoder first locally quantizes its observed source sequence and then sends a linear combination of the original source sequence and the quantized sequence over the channel. The receiver first decodes the transmitted quantized sequences and then combines them  with its channel outputs to estimate the two source sequences $S_1^n$ and $S_2^n$. Hybrid coding thus includes uncoded transmission and source-channel separation as special cases. For the MAC setup under consideration,  hybrid coding is the best scheme proposed so far and in fact, performs close to the fundamental limit of minimal distortions for all parameter values $\rho, \sigma^2, P, N$. 


Hybrid coding or other JSCC schemes have  been proposed for a variety of other multi-terminal networks, see \cite{Minero:IT:15, Padakandla2:IT:21, Bross:IT:16,Liu:IT:11, Murin:IT:14,  Gunduz:IT:07} for a few examples. \highlight{A canonical setup is the transmission of correlated sources over a broadcast channel (BC), for which both achievability \cite{Han:TIT:87a} and converse results \cite{Gohari:All:2008, Khezeli:TIT:15, Yu:TIT:18} have been proposed. Special attention was given for the transmission of correlated Gaussian sources over a Gaussian BC, for which  hybrid coding was first proved to generally improve over both uncoded transmission and SSCC \cite{Bross:TIT:2010}, and shortly thereafter shown to be optimal under a quadratic distortion constraint in all regimes of source correlations and source and channel noise variances  and for all admissible distortions \cite{Tian:TIT:11}.}

\begin{figure}
		\centering
		\includegraphics[width=3.5in]{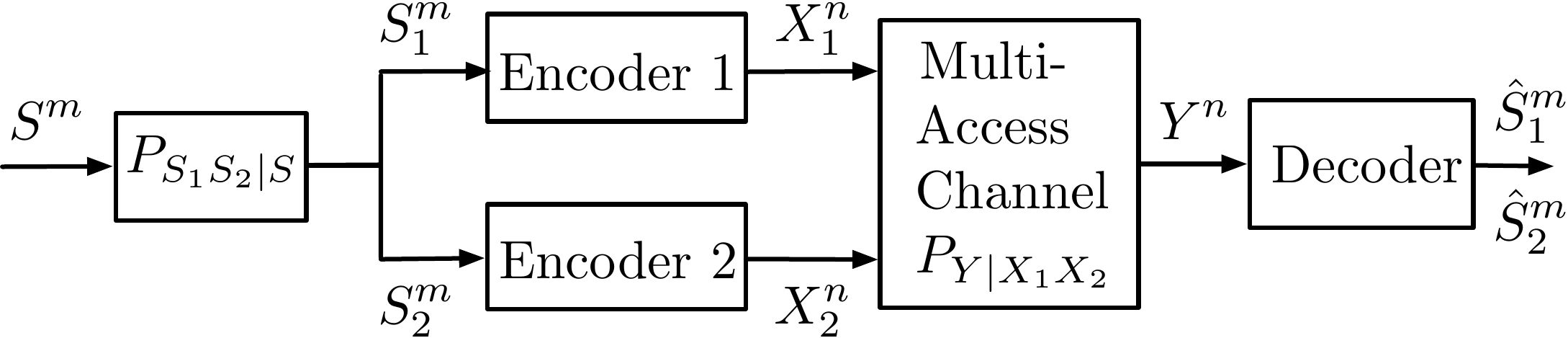}
		\caption{Joint CEO-MAC problem.}
  		\label{fig:CEO}
\end{figure}

\highlight{Related to the problem of sending correlated sources over a MAC is the joint CEO-MAC problem  in Fig.~\ref{fig:CEO}. Communication again takes place over a MAC, but the two encoders cannot directly observe the source that is of interest to the receiver. Instead, each encoder only observes a (correlated) noisy version of a hidden source $S^m$, which the decoder aims to reconstruct by the sequence $\hat{S}^m$. This setup was first introduced and studied in the special case in which the MAC degenerates to two rate-limited but noiseless links \cite{Berger:TIT:96, Viswanathan:TIT:97, Oohama:TIT:1998, Prabhakaran:ISIT:2004}. In this case the problem turns into a pure source-coding problem.  The variation with a Gaussian MAC was studied for example in \cite{Gastpar:IPSN:2003, Kochman:Allerton:2008}. The results in \cite{Gastpar:IPSN:2003} show that in some cases uncoded transmission of the sources is optimal. In general, however, JSCC schemes strictly improve over separation based architectures that combine an optimal CEO code with an optimal MAC code. The setup in which the source of interest cannot be directly observed at the encoders is of great interest  for semantic communication, as it models the desired feature (or class in a classification task) that is not directly observed at any of the terminals but can only be accessed indirectly through the data sample, e.g, an image.}

\highlight{Related to the joint CEO-MAC problem and to the problem of sending correlated sources over a MAC is also the problem of function computation over MACs \cite{Nazer:IT:2007}. In this problem the setup is again similar to the one in Fig.~\ref{fig:correlated_MAC}, except that the decoder wishes to reconstruct a symbolwise function of the two source sequences $f^{\otimes m} (S_1^m, S_2^m)$. As shown in \cite{Nazer:IT:2007}, lattice-based JSCC schemes outperform separation-based schemes in this scenario.}

\subsubsection{Broadcast channel (BC) with side information}

\begin{figure}
		\centering
		\includegraphics[width=3.5in]{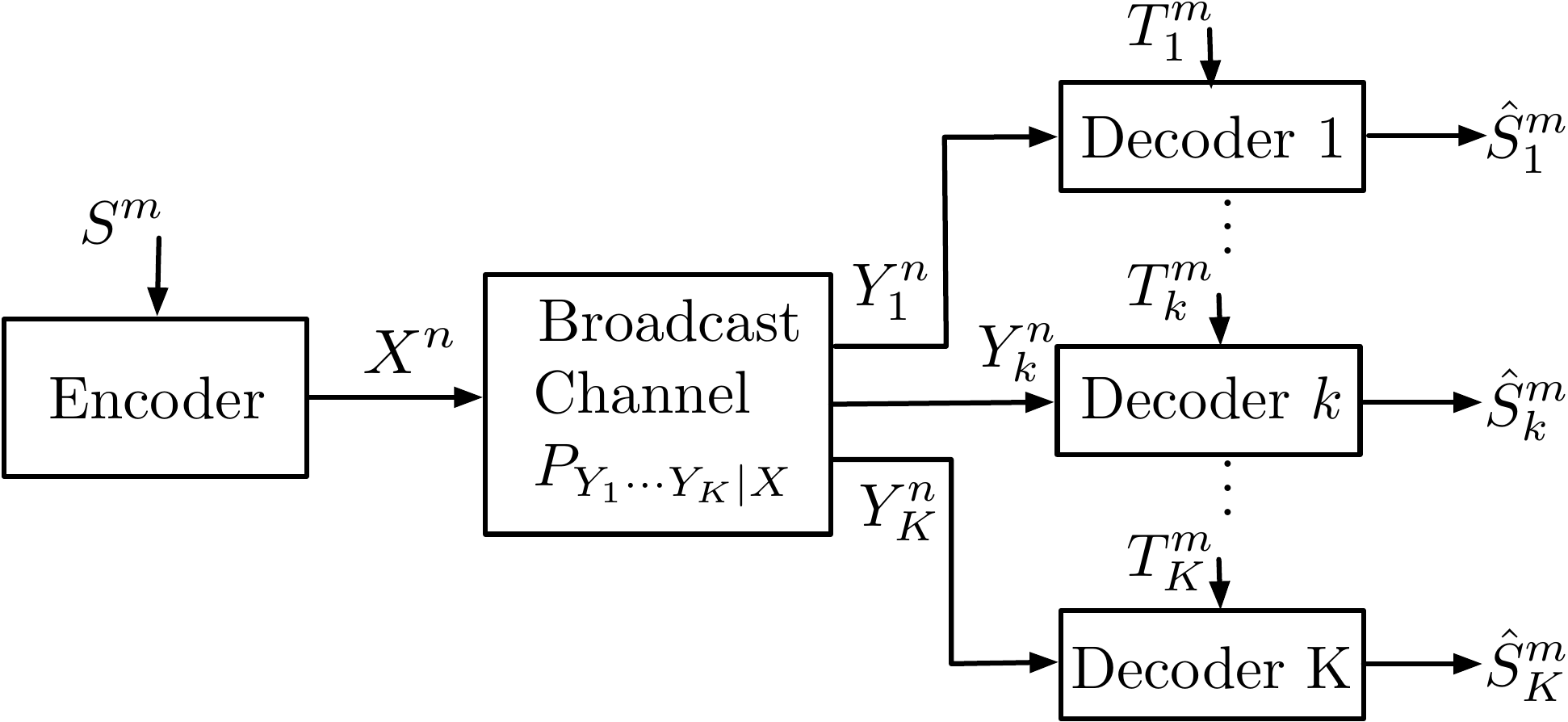}
		\caption{Slepian-Wolf coding over a broadcast channel (BC).}
  		\label{fig:SW}
\end{figure}

Let us first consider the generalization of the problem with side information, studied in Section \ref{ss:SW_JSCC}, to the broadcast channel scenario, as illustrated in Fig. \ref{fig:SW}. Here, the encoder transmits the outcome of a single source sequence $S^m$ to multiple receivers with side-information $T_1^n, \ldots, T_K^m$, respectively, over a discrete memoryless BC $P_{Y_1\cdots Y_K|X}$. This scenario is named as Slepian-Wolf coding over broadcast  channels (BC) by Tuncel in \cite{Tuncel:TIT:2006}.  


The generalized coding scheme introduced in Section \ref{ss:SW_JSCC} can be directly adopted in this scenario \cite{Tuncel:TIT:2006}. The encoder maps each source outcome to a different channel codeword, each of these codewords is generated in an i.i.d. fashion from a channel input distribution $p_X$. After observing the channel outputs $Y_k^n$ and the source side-information sequence $T_k^m$, decoder $k$ employs either the joint decoding or the list decoding scheme presented in Section \ref{ss:SW_JSCC}. 

It can be shown that the error probability, i.e., the probability that $\hat{S}_k^m\neq S^m$ tends to 0 as $n$ tends to infinity and $m=r\cdot n$, whenever 
\begin{equation}\label{eq:SW_cond}
  r \cdot   H(S|T_k) <  I(X;Y_k), \quad k=1, \ldots, K.
\end{equation}

Following the standard converse steps in the single-user Slepian-Wolf coding problem, it can be shown that the condition \eqref{eq:SW_cond} (with non-strict inequality) has to hold for some probability mass function (pmf) $P_X$. This leads to the following theorem, which is a slightly modified expression of Theorem 6 in \cite{Tuncel:TIT:2006}. 

\begin{theorem}\label{thm:SW_BC}
The source-channel coding capacity for the Slepian-Wolf coding problem over the broadcast channel is given by 
\begin{equation}
   \sup_{p(x)} \min_{k=1,\ldots, K} \frac{I(X;Y_k)}{H(S|S_k)}. \label{broadcast_cap}
\end{equation}
\end{theorem}

Above theorem shows that each user can compensate bad side-information with  good channel conditions and vice versa. For the reconstruction capability of a decoder only the global quality of the side-information and the channel matter, but not the individual qualities. In this problem, if we were to apply a SSCC approach, we would need to deliver $H(S|S_k)$ bits per source sample to decoder k. Hence, we need an achievable tuple $(R_1, \ldots, R_K)$ within the capacity region of the underlying broadcast channel which maximizes $R_k/H(S|S_k)$ over $k$. Tuncel showed in \cite{Tuncel:TIT:2006} through an example that this rate can be strictly below the source-channel capacity. 

In the described code construction, the encoder behaves like in SSCC. It is, however, important that the decoder jointly reconstructs the channel and source codewords. The proposed scheme is thus still a JSCC scheme, even if encoding is performed separately. An alternative coding scheme which applies SSCC that can still achieve the optimal source-channel capacity is also possible. In this alternative coding scheme, source samples are compressed separately for each user such that each user receives enough information to decode the source sequence when combined with its own side information sequence. Then, the source sequence is divided into blocks, and the bin indices for these blocks are transmitted to each receiver at a different channel block, and the users employ backward decoding. We will illustrate this coding scheme considering a network with two users. 

We fix $p(x)$ such that (\ref{eq:SW_cond}) holds. Then we consider a sequence of $(m, n)$ pairs such that $m/n$ is less than Eqn. (\ref{broadcast_cap}), and its limit is $r$, i.e., $\lim_{m,n \rightarrow \infty} m/n =r$. For each $(m,n)$ pair, a total of $Bm$ source samples will be transmitted over $(B+1)n$ channel uses. This corresponds to a source-channel code rate of $Bm/(B+1)n$, which becomes arbitrarily close to $m/n$ as $B \rightarrow \infty$.

\emph{Source code generation: } Corresponding to each decoder $i$, $i=1,2$, we consider $M_i = 2^{mR_i}$ bins, called the $\mathrm{T_i}$ bins. All possible source outcomes $s^m \in \mathcal{S}^m$ are partitioned randomly and uniformly into these bins, independently from each other, i.e., the distribution into $M_1$ bins is independent of the distribution into $M_2$ bins. The $i$th bin index of the source sequence represents the information that will be provided to user $i$. This bin assignment, which corresponds to source compression, is made available to all the users.

\emph{Channel code generation: } For the channel codebook, generate at random $M_1M_2$ channel codewords i.i.d. with probability $p(x^n(i,j))=\Pi_{t=1}^n p(x_{t})$, and index them as $x^n(j
i,j)$ with $(i,j) \in [1:M_1] \times [1:M_2]$.

\begin{figure}
\begin{tiny}
\begin{tabular}{|l|l|l|l|l|l|l|l|}
  \hline
    Block 1 & Block 2 & $\cdots$ & Block B & Block B+1 \\ \hline
    $x^n(w_{1,1}, 1)$ & $x^n(w_{2,1}, w_{1,2})$ & $\cdots$ & $x^n(w_{B,1}, w_{B-1,2})$ & $x^n(1, w_{B,2})$ \\ \hline
\end{tabular}
\caption{Channel codeword assignment for separation-based encoding and backward decoding scheme for Slepian-Wolf coding over BC with $K=2$ users. $B$ source blocks are transmitted in $(B+1)$ channel blocks.} \label{f:table_backward}
\end{tiny}
\end{figure}

\emph{Encoding: } Consider a source sequence $s^{Bm}$ of length
$Bm$. Partition this sequence into $B$ portions, $s_{b}^m$,
$b=1,\ldots,B$. Similarly, partition the side information sequences
into $B$ length-$m$ blocks $t_i^{Bm}=[t_{i,1}^m,\ldots,t_{i,B}^m]$ for $i=1,2$. The bin index of the $j$th block of the source output sequence $s_{j}^m$ with respect to $\mathrm{T_i}$ bins is denoted by $w_{j, i}$. The estimate of $w_{j,i}$ at user $k$, $k=1,2$, $k \neq j$, is denoted by $\hat{w}_{j,i}^k$. See Figure \ref{f:table_backward} for an illustration of the encoding scheme.

In block 1, transmitter observes $s_{1}^m$, and finds the corresponding $\mathrm{T_1}$ bin index $w_{1,1} \in [1:M_1]$. It transmits the channel codeword $x^n(w_{1,1},1)$. In block $2$, it transmits the channel codeword $x^n(w_{2,1}, w_{1,2})$ corresponding to the $\mathrm{T_1}$ bin index of the second source block and the $\mathrm{T_2}$ bin index of the first source block. In the following blocks $b=2,\ldots,B$, the it transmits the channel codewords $x^n(w_{b,1}, w_{b-1,2})$ where $w_{b,i} \in [1:M_i]$ for $i=1,2$. In block $B+1$, it transmits $x^n(1, w_{B,2})$.

The first user estimates the source block $s_{b-1}^m$ at the end of block $b-1$, denoted by $\hat{s}_{1,b-1}^m$, and finds the corresponding bin index $\hat{w}_{b-1, 2}^1 \in [1:M_2]$. When the condition in \eqref{eq:SW_cond} holds, it is possible to show that the user will recover each source block successfully with vanishing error probability as $m, n \rightarrow \infty$. 

On the other hand, user 2 will wait until the end of last block to start decoding, and will decode the source blocks backwards. At the end of channel block $B+1$, user $2$ decodes the first source block $s_{1}^m$, denoted by $\hat{s}_{2,B}^m$, and finds the corresponding bin index $\hat{w}_{B, 1}^2 \in [1:M_1]$. Having estimated this bin index, it can then decode the next source block $B-1$, and so on so forth. Similarly, each source block will be decoded by the second user with high probability if \eqref{eq:SW_cond} holds.



The idea of Slepian-Wolf coding over broadcast channels is extended to lossy broadcasting scenario in \cite{Nayak:TIT:10}. This coding principle is also very useful in cache-aided networks, where it was termed joint cache-channel coding \cite{Saeedi:IT:18, Amiri:TCOM:18, Saeedi:IT:19}. In cache-aided networks users store contents at cache memories close to end users, which then serves as side-information during the actual communication phase. The situation is thus similar to the one described above, but we face a data communication problem where different decoders wish to learn different parts of the data.

\subsection{Types of source-channel separation}\label{ss:SCS}

Above results and coding schemes beg further discussion on source and channel separation in multi-user networks. In \cite{Gunduz:TIT:09}, two types of separations are proposed: The first type of separation is a direct generalization of Shannon's Separation Theorem to multi-user networks, and answers the question ``Can we determine the source-channel capacity by simply comparing the source coding rate region with the capacity region of the underlying sources and channels?'' This is trivially the case in point-to-point channels. This also holds for a MAC when there is no multi-user interference, that is, the users have orthogonal channels to the receiver \cite{Xiao:TIT:07}. In such a case, the source-channel capacity for a given target distortion tuple $(D_1, \ldots, D_K)$ and input cost constraint tuple $(P_1, \ldots, P_K)$ is given by $\sup_{(R^s_1, \ldots, R^s_K) \in \mathcal{R}_{comp}, (R^c_1, \ldots, R^c_K) \in \mathcal{R}_{chan}} \min_{k=1, \ldots, K} R^c_k / R^s_k$,
where $\mathcal{R}_{comp}$ denotes the source coding rate region for the considered sources for target distortion tuple $(D_1, \ldots, D_K)$, and $\mathcal{R}_{chan}$ denotes the capacity region of the underlying channel with cost constraints $(P_1, \ldots, P_K)$.

Note that the two regions are characterized completely independently of each other; that is, the optimal source coding rate region is ignorant of the channels, and the channel capacity region is ignorant of the source statistics. Moreover, the coding schemes that achieve the operating points in these two rate regions are also oblivious of each other, apart from exchanging bits. On the other hand, the coding scheme achieving the optimal source-channel capacity in Theorem \ref{thm:SW_BC} points to a different type of separation. The source-channel capacity can still be determined by comparing two rate regions that are statistically independent of each other; however, the coding schemes to achieve these rate regions are not oblivious to each other, and moreover, the channel coding rate region does not correspond to the capacity region of the underlying channel. In \cite{Tuncel:TIT:2006}, these two types of separations are called \textit{`informational separation'} and \textit{`operational separation'}, respectively.

\begin{figure}
		\centering
		\includegraphics[width=3.5in]{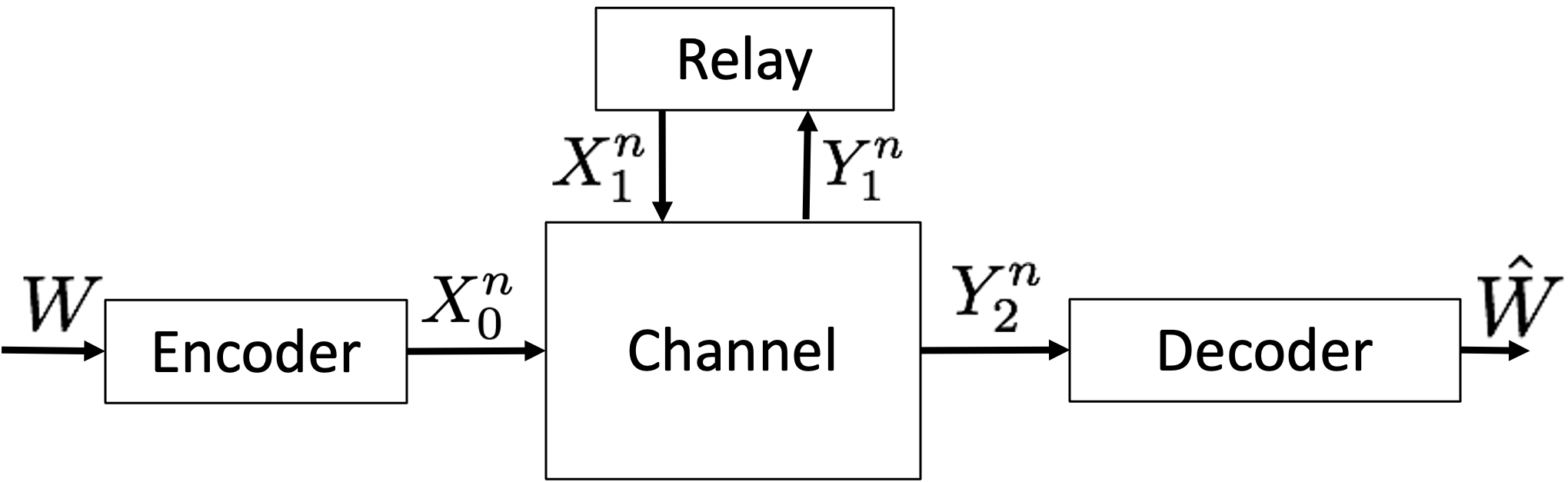}
		\caption{The relay channel.}
  		\label{fig:relay_channel}
\end{figure}

\subsection{JSCC for channel coding}\label{ss:JSCC4CC}

So far, we have focused on JSCC for transmitting source signals to be recovered at the decoder under a prescribed distortion measure. However, JSCC in general can also be instrumental in increasing the achievable coding rates in multi-terminal networks. Treating received signals as sources to be forwarded to other terminals to help their decoding of the underlying message was first considered in \cite{Cover:TIT:79} in the context of a relay channel (illustrated in Fig. \ref{fig:relay_channel}). Here, the goal of the encoder is to convey message $W \in \{1, \ldots, 2^{nR}\}$ to the decoder. The relay terminal can overhear the transmission, and would like to help the decoder by conveying extra information about the transmitted message. One option is for the relay to first decode $W$ and forward it to the decoder in collaboration with the encoder. However, if the relay's channel is not strictly better than that of the decoder, requiring correct decoding at the relay can become a bottleneck instead. An alternative scheme is to treat the signal received at the relay, $Y_1^n$, as a source signal, which is to be forwarded to the decoder with as high fidelity as possible. Cover and El Gamal treated this problem as a source and channel coding problem. Since it is a point to point channel, separation is known to be optimal in this case, and hence, the relay can first compress $Y_1^n$ and forward it to the decoder using channel coding. Note, however, that, the decoder's received signal, $Y_2^n$ is correlated with that of the relay, and hence, can act as correlated side information, as in Wyner-Ziv coding. Let $\hat{Y}_1^n$ denote the compressed version of relay's received signal. To forward it to the decoder, the following condition should hold: $I(Y_1; \hat{Y}_1|X_1, Y_2) \leq I(X_1; Y_2)$. Then the decoder can combine the compressed version of relay's signal with its own received signal to decode the message transmitted by the encoder. The achievable rate with this compressed-and-forward scheme is then given by $I(X_0; \hat{Y}_1, Y_2|X_1)$, where the joint distribution of the random variables is given by $p(x_0)p(x_1)p(\hat{y}_1|x_1, y_1)p(y_1, y_2|x_0, x_1)$.

In the case of multi-user relay channels, the relay will need to broadcast its received signal to multiple receivers for an extension of the compress-and-forward scheme. Note, however, that there is no separation theorem in this case since the goal is to send a lossy version of the relay's signal to multiple receivers each with a different correlated side information (similarly to \cite{Nayak:TIT:10}. The simplest scenario requiring the broadcast of relay signal to multiple users would be the separated two-way relay channel model. In this model, two users try to communicate with each other with the help of a relay terminal. It is called separated since there are no direct links between the users, and hence, they rely on the relay terminal for communications. The relay receives the superposition of the signals transmitted by each user. In general, it may not be feasible for the relay to decode the users' messages, and instead it can broadcast its received signal to both users. Since each user already knows its own transmitted signal, it may be easier for them to cancel their own interference and decode the other user's message. In this case, user's own signals act as correlated side information when the relay's received signal is broadcast to the users. Two achievable rate regions are proposed in \cite{Gunduz:Allerton:08} relying on JSCC of relay's signal using Wyner-Ziv coding over the broadcast channel from the relay to the users. A single compressed version of the relay's signal is transmitted to both users in the first scheme, while a refinement layer is sent to one of the users in the second scheme. Further gains are possible by introducing a hybrid relaying scheme, which exploits amplify-and-forward relaying together with the proposed digital broadcasting approaches, similarly to \cite{Gunduz:ISIT:08}. While these schemes generally lead to achievable rates or rate regions, and are not capacity achieving, they are shown to achieve the optimal diversity-multiplexing trade-off for MIMO relay and two-way relay channels in \cite{Yuksel:TIT:07} and \cite{Gunduz:Asilomar:08}, respectively.

\begin{figure}
		\centering
		\includegraphics[width=3.5in]{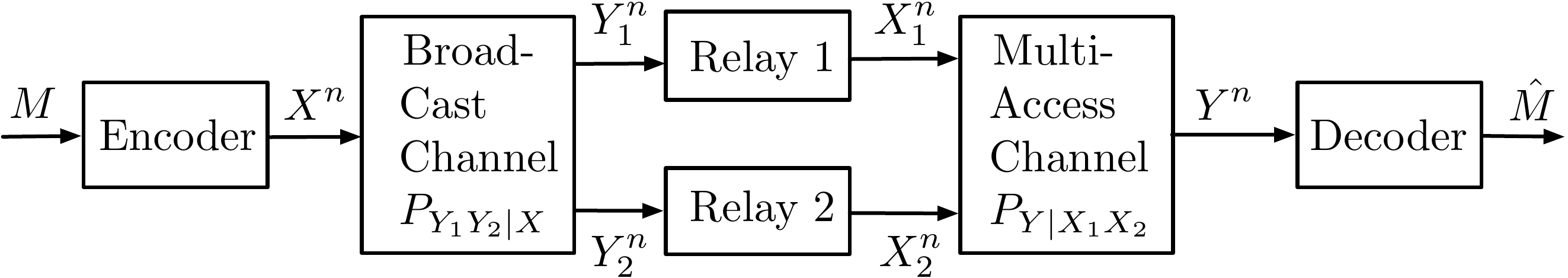}
		\caption{Data communication over the diamond  relay channel. }
  		\label{fig:diamond}
\end{figure}

Hybrid coding (which we used to send correlated sources over a  MAC) can also improve the performance of data communication problems.  Consider, for example, the diamond relay network in Fig.~\ref{fig:diamond}, where the goal is to send a message from the encoder to the decoder with the help of two relays. The  best known coding scheme (in terms of achievable rates) uses hybrid coding at the relays, i.e., the relays quantize their channel outputs $Y_1^n$ or $Y_2^n$ into source codewords $U_1^n$ or $U_2^n$, and then send a symbolwise function of these source codewords and the observed channel outputs. Similarly to the MAC with correlated sources, hybrid coding allows to transfer part of the correlation between the outputs $Y_1^n$ and $Y_2^n$ to the inputs $X_1^n$ and $X_2^n$, which for certain MACs allows for a better communication performance than using independent inputs.

JSCC plays a prominent role also in interactive data communication scenarios when terminals wish to communicate messages and not source sequences. In fact, interactive data communication schemes often (e.g., \cite{Venkataraman:TIT:11, VenkataramananBC:IT:2013, Wu:IT:16, Shannon1961TwowayCC, Han:TIT:89, Han:IT:84}) build on  strategies where information is transmitted in a given block and refined in subsequent blocks. From a receiver's point of view, during the refinement blocks  the data bits are no longer independent  bit sequences   because they are   correlated with the symbols observed during the previous blocks. In this sense, the communication problem  turns into a JSCC  problem with side-information, and JSCC generally  outperforms separation-based schemes. 

The refinement idea is for example exploited in \cite{Wu:IT:16} on a two-user broadcast channel (BC) with rate-limited feedback, i.e., on a single-transmitter two-receiver channel where both receivers can send back feedback information over rate-limited communication links. A simple but very efficient scheme \cite{Wu:IT:16} for this setup is that, after each communication block, the receiver with the worse channel condition quantizes its observed output signal and sends back the quantization information over the rate-limited link. In the next block this information is sent from the single transmitter to the receiver with the better channel as part of the \emph{cloud center}, i.e., as part of the information that has to be decoded by both receivers. The weaker receiver will not be bothered by this additional transmission in the cloud center since it already knows the quantization information (it generated itself), and can simply ignore its presence. The stronger  receiver has sufficiently good channel conditions to be able to decode the quantization information and  reconstruct a quantized version of the  signal observed at the weaker receiver in the previous block. This second output signal allows the stronger receiver to decode the previous block with a single-input multi-output (SIMO) decoding based on two outputs. On an abstract level, the described scheme establishes a communication path  (over the feedback and forward links) from the weaker receiver to the stronger receiver, which allows to exchange information about the output signals essentially without using resources on the forward link. 

The described scheme is a JSCC scheme because it does not treat the quantization information as an isolated sequence of bits, but instead, takes into account that it is information obtained from the previous' block's output at the weak receiver, and thus known to it. Without taking into account this knowledge, the proposed scheme does not yield any benefit compared to a simple non-feedback scheme. However, after accounting for this side-information the scheme improves over the optimal no-feedback scheme on a large class of channels as long as the the feedback rate is non-zero \cite{Wu:IT:16}. Incidentally, JSCC schemes that improve capacity with feedback for Gaussian BCs with noisy feedback are proposed in \cite{Ravi:IT:21}, which leads to a complete characterization when capacity improvements are possible and when not. 

 \begin{figure}
		\centering
		\includegraphics[width=3in]{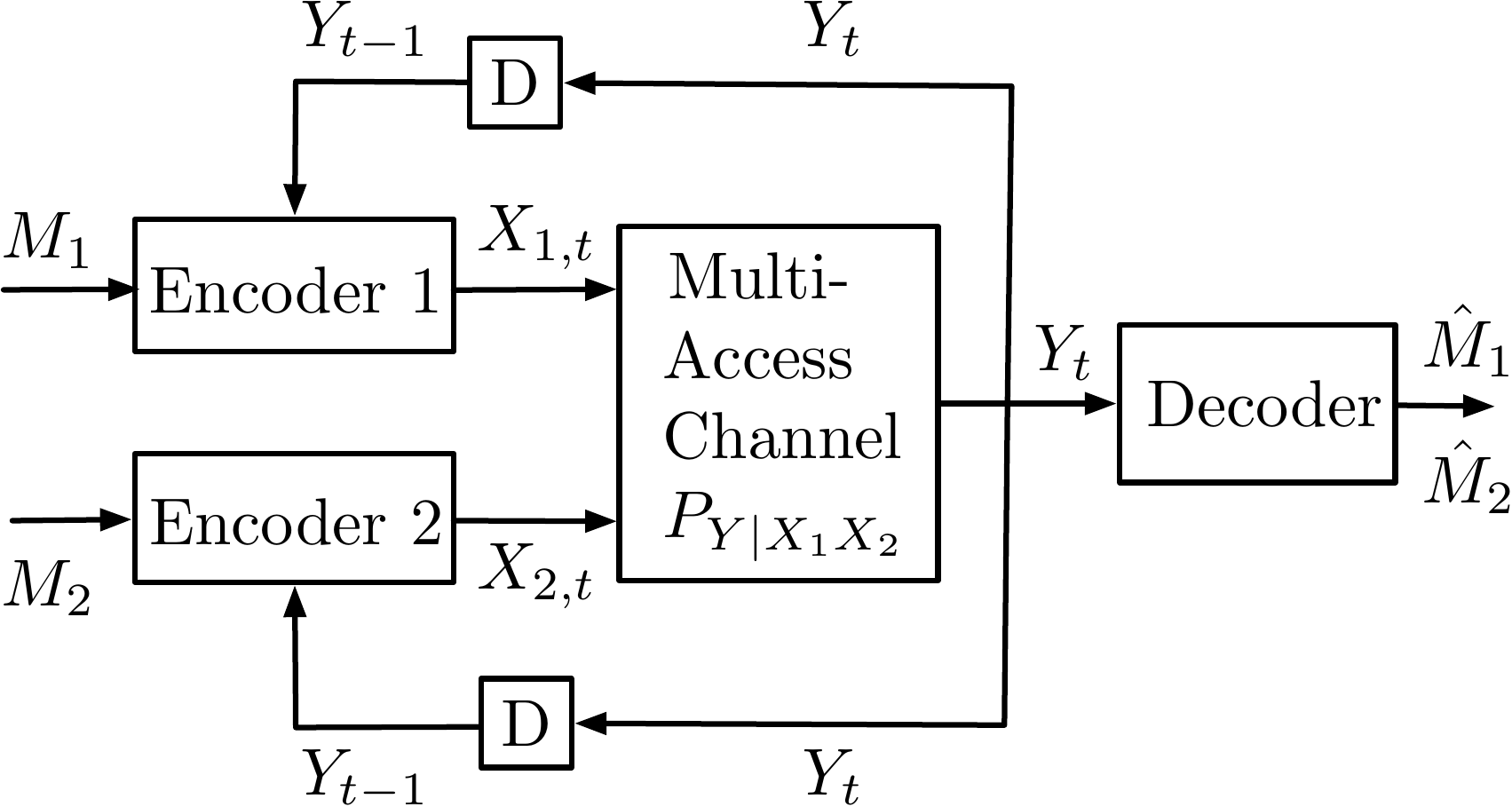}

		\caption{Data communication over the two-user MAC with feedback.}
  		\label{fig:MAC_fb}
	\end{figure}

Another canonical instance of interactive communication is the MAC with feedback problem in Fig.~\ref{fig:MAC_fb}, 
where two encoders communicate to a single decoder sequentially, in the sense that the time-$t$ channel inputs $X_{1,t}$ and $X_{2,t}$ are functions of the respective messages $M_1$ or $M_2$, as well as the past channel outputs $Y^{t-1}$. For this setup, a block-Markov coding scheme is proposed in \cite{Venkataraman:TIT:11} , where in each block JSCC is used to send refinement information about the previously transmitted codewords. This refinement information is calculated based on the observed feedback outputs, and can be highly correlated across the two transmitters. Sending such correlated side-information in the next block allows to correlate the channel inputs from the two transmit terminals (without resorting to trivial inputs) and  to achieve higher communication rates.

An elegant coding scheme for the two-user multi-access channel with feedback in the Gaussian case was proposed in  \cite{Ozarow:TIT:84}.  Each of the two transmitters first maps its message $M_k$, $k\in\{1,2\}$ to one of $2^{nR_k}$  equally-spaced points on a grid over the interval $[-1/2, 1/2)$, i.e., 
\begin{equation}
    \Theta_k = \frac{M_k}{2^{nR_k}} - \frac{1}{2}
\end{equation}
and then sends this message point during the first channel use by transmitting $X_{k,1}= \sqrt{12 P_k} \Theta_k$, where the scaling is performed to satisfy the input power constraint $P_k$. In subsequent channel uses, Tx~$k\in\{1,2\}$ calculates the linear minimum mean squared error (LMMSE) $\hat{\Theta}_{k,t}$ of $\Theta_k$ based on past feedback outputs $Y_{1}, \ldots, Y_{t-1}$ and sends a scaled version of the LMMSE error $\epsilon_{k,t}$, as in the single-user scheme explained in Subsection~\ref{ss:JSSC_Feedback}. Following again the idea in Subsection~\ref{ss:JSSC_Feedback}, the two transmitters then send a scaled version of their estimation errors so as to satisfy the power constraint, while one of the two transmitters also modulates its inputs with a $\pm 1$ sequence to ensure that at each time-instance the inputs of the two transmitters are positively correlated and thus profit from each other. The receiver calculates the LMMSE estimates $\hat{\Theta}_{1,n}$ and $\hat{\Theta}_{2,n}$ of both message points and declares the messages  $M_1$ and $M_2$ associated to the grid points that are closest to these estimates. 

The described scheme is not only particularly simple to implement but achieves the  sum-capacity of the two-user Gaussian MAC with output feedback. It has been extended to MACs with more than two users \cite{Kramer:TIT:02}, for which it was shown to achieve capacity \cite{Kramer:TIT:02, Sula:IT:20}, to interference and broadcast channels \cite{Ozarow:IT:85, Ardestanizadeh:IT:12, Murin:IT:14,Murin:SPL:15}, and to MACs and BCs with   noisy feedback \cite{Lapidoth:IT:10, Ravi:IT:21, Tian:TIT:17}



\subsection{Beyond average distortion}\label{ss:beyond_average}

The classical JSCC problem considers  an average  distortion measure, see \eqref{eq:add_Dist}, where each summand  depends only on the pair of source and reconstruction symbols at a given time $i$, which is key to the proof of Shannon's Separation Theorem. Depending on the application, however, more general performance criteria have to be applied. 

\begin{figure}[t]
	\centering
	\includegraphics[scale=0.24]{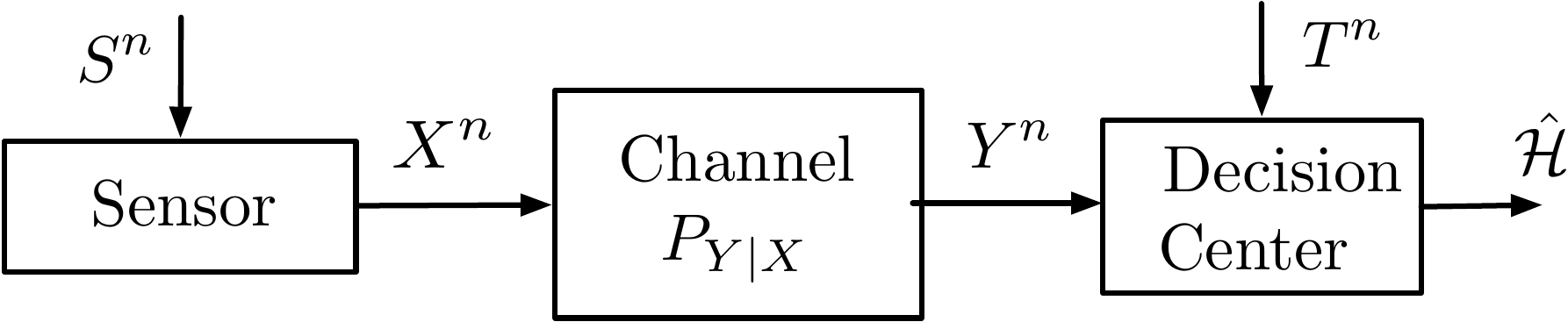}
	\caption{Distributed hypothesis testing scenario.}
	\label{fig:DHT}
\end{figure}

Consider, for example,  the distributed hypothesis testing problem in Fig.~\ref{fig:DHT} without bandwidth mismatch, i.e., $m=n$, where a sensor (transmitter) observes  source sequence $S^n$ and communicates to a decision center over a noisy communication channel. Based on the outputs of the channel and its own observation $T^n$ the receiver  has to guess the joint distribution that underlies the distributed observations $S^n$ and $T^n$. A binary hypothesis is assumed where under the null hypothesis:
\begin{equation}
\mathcal{H}=0\colon \quad  (S^n,T^n)\quad \text{i.i.d.}\; \sim P_{ST},
\end{equation}
and under the alternative hypothesis:
\begin{equation}
\mathcal{H}=1\colon \quad (S^n,T^n)\quad \text{i.i.d.}\; \sim Q_{ST}.
\end{equation}
for two given pmfs $P_{ST}$ and $Q_{ST}$. 

Mathematically speaking, the sensor maps the observed source sequence  $s^n$ to channel inputs $x^n(s^n)$  and the decision center maps the pair of channel outputs $y^n$ and observed side-information $t^n$ to a decision $\hat{\mathcal{H}}(y^n,t^n)$.

An exponent $\theta$ is said achievable, if for  sufficiently large blocklengths $n$, there exist encoding and decoding functions  such that the decision center's  probabilities of error 
	\begin{eqnarray}
	\alpha_{n}&:= \Pr[\hat{\mathcal{H}}=1|\mathcal{H}=0],
	\\
	\beta_{n}&:= \Pr[\hat{\mathcal{H}}=0|\mathcal{H}=1],
	\end{eqnarray}
	satisfy 
	\begin{equation}
	   \lim_{n\to \infty}\alpha_{n} =0 ,\label{def1bb}
	\end{equation}
	and 
	\begin{equation}
	   -\varlimsup_{n\to \infty}\frac{1}{n}\log_2\beta_{n} \geq \theta. \label{def2bb}
	\end{equation} 
The goal  is to maximize the achievable type-II error exponent $\theta$.

For certain families of source distributions $P_{ST}$ and $Q_{ST}$, for example, for testing against independence, where $Q_{ST}=P_{S}P_T$, a  separation-based scheme \cite{Saleh:IZS:18} achieves the optimal error exponent $\theta$. In particular, for testing against independence the optimal exponent is $\theta^\star = \max I(U;T)$, where the maximization is over all auxiliary random variables $U$ such that the tuple $(U, S,T)\sim P_{U|S}P_{ST}$ satisfies $I(U;T) \leq C$, with $C$ denoting the capacity of the communication channel as before. 

\begin{figure*}[t]
	\centering
	\includegraphics[scale=0.25]{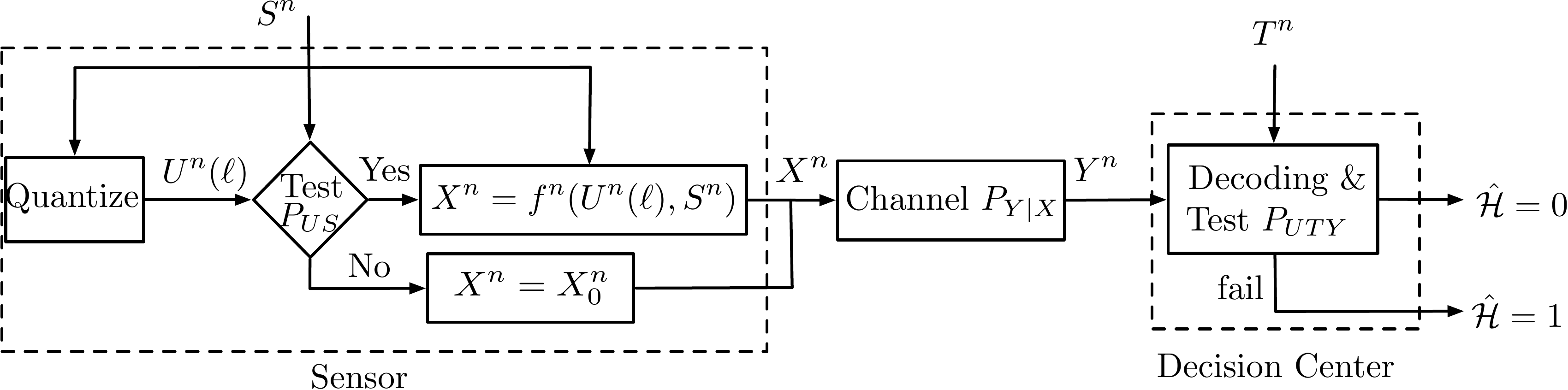}
	\caption{A distributed detection scheme based on JSCC (hybrid coding) and UEP.}
	\label{fig:coding}
\end{figure*}

In general, however, an improved performance can be attained by a JSCC scheme. The best JSCC schemes known to date are  the schemes in  \cite{Sreekumar:IT:20, Saleh:IT:20}, which combine either hybrid coding or joint decoding \`a la Tuncel \cite{Tuncel:TIT:2006} with unequal error protection (UEP) \cite{Borade:IT:09}. A slightly simplified version of the  coding scheme in \cite{Sreekumar:IT:20} based on hybrid coding is sketched in Fig.~\ref{fig:coding} and described in the following. The full scheme in \cite{Sreekumar:IT:20} is obtained by additionally introducing coded time-sharing and dependence between various code components. \\[-0.1ex]

We pick a conditional pmf $P_{U|S}$ and a function $f$ mapping $(U,S)$ to the channel input $X$. The joint pmf is given by $P_{USTXY}=P_{U|S}P_{ST}P_{X|US}P_{Y|X}$, where $P_{Y|X}$ indicates the channel law and $P_{X|US}$ the input distribution obtained when applying the chosen function $f$ to $(U,S)$. We assume that $I(U;S)>R> I(U;S)-I(U;T)=I(U;S|T)$  with respect to the joint pmf $P_{UST}$. 
We then generate a random codebook $\big\{u^n(\ell)\colon \ell=1, \ldots, 2^{nR'}\big\}$, by picking all entries i.i.d. according to the marginal $P_{U}$. \\[-0.1ex]

\noindent\underline{\textit{Sensor:}} Given that it observes the source sequence $S^n=s^n$, the remote sensor looks for an index $\ell$ so that the codeword $u^n(\ell)$ is jointly typical with the observed source sequence $s^n$ according to the joint pmf $P_{US}$. It then applies the function $f$  componentwise to the chosen codeword $u^n(\ell)$ and  the source sequence $s^n$, i.e., $x^n=f^{n}(u^n(\ell), s^n)$, and sends the result over the channel.   If no typical codeword can be found, then the sensor  sends a special input sequence $x_0^n$ over the channel. In other words, if the typicality test succeeds, the hybrid encoding steps from \cite{Minero:IT:15} are used, and otherwise, the UEP mechanism from \cite{Borade:IT:09} is employed. 

The idea is to choose the sequence $x_0^n$  to be well distinguishable at the receiver from any of the codewords $u^n(\ell)$, thus implementing an UEP code. This  is crucial for the present hypothesis testing scenario, because it allows to specially protect the message indicating that none of the codewords has an empirical statistics close to $P_{US}$, which is the statistics that we can expect under $\mathcal{H}=0$ for at least one of the codewords if we choose the codebook of the indicated size. The absence of such a codeword thus hints to the hypothesis $\mathcal{H}=1$ and the sensor applies a UEP code to enforce high reliability on the transmission of this information, because in our asymmetric setup we have more stringent requirement on missing hypothesis $\mathcal{H}=1$ (we require exponentially decreasing $\beta_n$) than on missing hypothesis $\mathcal{H}=0$ ($\alpha_n$ can decay arbitrarily slowly.\\[-0.1ex]

\noindent\underline{\textit{Decision Center:}} Assume that  $T^n=t^n$ and $Y^n=y^n$. The receiver first looks for an index $\ell'\in \{1,\ldots,\lfloor 2^{nR} \rfloor\}$ such that  the following two conditions are simultaneously satisfied:
\begin{equation}\label{eq:channel_dec}
(u^n(\ell'),y^n,t^n) \in \mathcal{T}_{\mu}^n(P_{UYT}),
\end{equation}
and
\begin{equation}
H_{\text{tp}(u^n(\ell'),t^n)}(U|T) = \min_{\tilde{\ell}} H_{\text{tp}(u^n(\tilde{\ell}),y^n)}(U|T),\label{MI}
\end{equation}
where here $\mathcal{T}_{\mu}^n(P_{UYT})$ denotes the set of triples $(u^n,y^n,t^n)$ with empirical statistics equal to $P_{UYT}$ within a small positive $\mu>0$ margin, and $H_{\text{tp}(u^n,y^n)}(U|T)$ denotes the conditional entropy of  random variables $U$ given $T$ whose joint pmf is given by the empirical statistics (the type) of the pair $(u^n, y^n)$. If the tests in \eqref{eq:channel_dec} and \eqref{MI} are simultaneously successful for the same index $\ell'$,  then the decision center declares $\hat{\mathcal{H}}=0$, and otherwise $\hat{\mathcal{H}}=1$.

The empirical conditional entropy test in \eqref{MI} is inspired by the distributed hypothesis testing scheme over a noiseless communication channel in \cite{Shimokawa:ISIT:94}. Recently, an improved test for the noiseless setup has been proposed \cite{Kochman:Arxiv:23}, and it can be expected that using their scheme in this setup with a noisy communication channel will also lead to an improved performance.

\highlight{The study of distributed hypothesis testing over noiseless links has received significant attention in the literature, with several remarkable results \cite{Han:TIT:87b, Han-Amari-1989,Ahlswede-Csiszar,Saleh:IT:20, HanAmari:IT:95, Shalaby:TIT:92, Berger:IT:79, WK-2017,Watanabe:ISIT:2022}. More complicated setups over noisy networks have  been considered for example in \cite{Saleh:Globecom:2018,Saleh:TCOM:2018, Saleh:JSAIT:2020}.}

\section{Practical Code Design for JSCC: Classical Approaches}\label{s:codes_classical}

Design of practical JSCC schemes has been a longstanding challenge in information and coding theory. The efforts can be grouped into three general categories. In the first group are joint code designs for idealistic i.i.d. or Markov source distributions. These approaches aim at extending some of the known code designs to exploit the redundancy in the source distribution. In the second group are joint designs for practical sources, such as images or videos, which typically consider two separate codes for compression and channel coding, whose parameters are optimized jointly. The last group are closer to a fully joint design, inspired by the optimality of uncoded/linear transmission for Gaussian sources over Gaussian channels, and employs linear encoding schemes with some hand-designed non-linearities for the transmission of practical image or video sources. Next, we present overviews of codes under these three categories.

\highlight{
\subsection{Practical JSCC Code Design}
There have been many practical coding design for JSCC based on the classical coding schemes, such as Turbo, low-density parity-check (LDPC), and polar codes. In his 1948 paper \cite{Shannon}, Shannon already wrote that ``... any redundancy in the source will usually help if it is utilized at the receiving point. In particular, if the source already has redundancy and no attempt is made to eliminate it in matching to the channel, this redundancy will help combat noise.'' He also provided a simple example in \cite{Shannon_RD} showing that uncoded transmission of a non-uniform binary source over a binary erasure channel can meet the optimal Shannon lower bound achieved by SSCC with infinite blocklength. Various designs have been proposed for general source and channel distributions to benefit from the source redundancy. Optimality of linear codes for lossless coding was observed in \cite{Hellman:TIT:75}. It was also shown in \cite{Ancheta} that linear codes cannot achieve optimal performance when lossy transmission is considered. Most follow up studies on explicit code construction focused on lossless tranmission. In \cite{hagenauer1995source}, Hagenauer proposed a modified version of the Viterbi algorithm that can benefit from the \textit{a priori} or \textit{a posteriori} information about the source bits. A joint trellis representation of the decoder is proposed in \cite{Murad:ITW:98} for i.i.d. and Markov sources that can allow Viterbi-like joint decoding of variable-length source codes with convolutional and trellis channel codes. Another trellis representation is suggested in \cite{Hagenauer:ITW:01} for variable-length Huffman coding combined with a convolutional code, and the Turbo principle is applied at the decoder to iterate between the source and channel decoders. The idea to exploit the residual redundancy of the compression scheme is also considered in \cite{Sayood:TCOM:20} and \cite{Lonardi:TIT:07}. The former combines differential pulse code modulation (DPCM) with a non-binary convolutional channel code and employs a Viterbi decoder. In the latter, Lempel-Ziv coding is shown to correct a few errors over the channel by exploiting the redundant bits that remain after compression. Residual redundancy in Huffman coding is exploited in \cite{Jeanne:TCOM:05} when transmitting a Markov source over a noisy channel, with both a Viterbi and a Turbo channel decoder. A first-order binary Markov source is considered in \cite{Zhu:TWC:06} with a Turbo channel code, and both constituent decoders are modified to optimize the performance. JSCC of more general hidden Markov source models is studied in \cite{Garcia-Frias:CL:97, Garcia-Frias:JSAC:01} with Turbo decoders, where a universal Turbo decoder is used in the latter, which means that the source statistics need not be known in advance.}

\highlight{
JSCC of a first-order Markov source using an LDPC code is considered in \cite{Kfir:ICECS:04}. A convergence analysis and a general method for the optimisation of irregular LDPC codes for joint source-channel receivers are presented in \cite{Poulliat:CL:05}. A lossless JSCC scheme based on two concatenated LDPC codes is proposed in \cite{fresia2009optimized}. An LDPC-based source encoder is concatenated with an LDPC channel encoder at the transmitter side, while a single joint belief propagation (BP) decoder is employed at the receiver. The asymptotic performance of the proposed scheme is theoretically analyzed using extrinsic information transfer (EXIT) charts. To reduce the implementation complexity of the JSCC system and to improve its performance in the waterfall region, double protograph LDPC (DP-LDPC) codes are proposed in \cite{He:ISCIT:12}. In general, the optimal channel code designed for the underlying channel is not optimal in the JSCC context due to the joint decoding operation. A joint protograph extrinsic information transfer (JPEXIT) analysis is proposed in \cite{Chen:CL:16} to calculate the decoding threshold of the joint protograph matrix, and the channel code in the DP-LDPC system is redesigned to improve the overall performance in the waterfall region. Further variants of DP-LDPC codes have been introduced to improve the performance \cite{Chen:CL:18, Chen:CL:19, Xu:CL:21}.  A graph theoretic construction method is proposed in \cite{Lv:Entropy:23} to transform the parity-check matrices of the source and channel codes into an inter-set constraint problem. Spatially coupled LDPC (SC-LDPC) codes are proposed in \cite{golmohammadi2021concatenated} to implement JSCC by concatenating two SC-LDPC coding modules for source and channel coding, respectively, at the transmitter with a joint belief propagation (BP)-based decoder. A joint source and channel anytime coding scheme with partial joint extending window decoding is proposed in \cite{Deng:TCOM:23} based on spatially coupled repeat-accumulate (SC-RA) codes. It is shown that the proposed approach provides low latency and high reliability transmission with respect to \cite{golmohammadi2021concatenated} at the cost of higher decoding complexity. 
}

\highlight{
A double polar code (D-Polar)-based scheme is proposed to implement JSCC in \cite{dong2021joint}, where the source compression is implemented by a polar encoder before the channel error correction is performed by a systematic polar code. 
At the receiver side, a turbo-like BP channel and source decoders are introduced to perform channel and source decoding, respectively. 
A joint BP decoding approach is instead applied in \cite{Dong:Electronics:22}, together with a biased extrinsic information transfer (B-EXIT) convergence analysis. 
}

\highlight{
Rateless codes, such as Raptor codes, have also been applied to implement JSCC\cite{bursalioglu2013joint}.
In particular, the rateless property of Raptor encoders can be  utilized during the explicit entropy coding stage to reduce the channel post-decoding residual errors. 
}

\highlight{
For lossy JSCC of memoryless sources, Kurtenbach and Wintz studied the problem of optimum quantizer design when the quantizer's output is transmitted over a noisy channel \cite{Kurtenbach:TCT:69}.  This is further studied in \cite{Farvardin:TIT:87} considering jointly the code assignment problem. Farvardin and Vaishampayan later extended their study to vector quantizers in \cite{Farvardin:IT:90, Farvardin:IT:91}. In \cite{Farvardin:IT:90}, Farvardin employed simulated annealing for the codeword assignment problem, and showed that when some statistical information about the channel is available, it is possible to improve the performance of the vector quantizer through a joint design of the codebook and the binary codeword assignment.
}

In \cite{Dunham:TIT:81}, Dunham and Gray proved that for stationary and ergodic sources, there exist optimal joint source and channel trellis encoding systems over discrete memoryless channels, which consist of concatenating a source tree coding system with a channel tree coding system, and perform arbitrarily close to the distortion rate function evaluated at the capacity of the underlying channel. While this is an existence result, in \cite{Ayanoglu:TIT:87}, Ayanoglu and Gray applied the generalized Lloyd algorithm to design joint source and channel trellis codes. Experimental results on independent and autoregressive Gaussian sources over binary symmetric channels under absolute and squared error distortion measures indicate that the joint system can outperform a separately optimized tandem system. A JSCC scheme using combined trellis coded quantization (TCQ) and trellis coded modulation (TCM) is proposed in \cite{Fischer:TCOM:91} for communication over an additive white Gaussian noise (AWGN) channel. Trellis-coded quantization (TCQ) of memoryless sources is developed in \cite{Wang:TIT:94} for transmission over a binary symmetric channel. It is shown to achieve the same performance as the one proposed in \cite{Ayanoglu:TIT:87}, but with a much lower complexity. Joint design of trellis-coded quantization/modulation (TCQ/TCM) is also proposed, and is shown to outperform cascade of separately designed TCQ and TCM systems, particularly at low SNRs. Further improvements in moderate to high SNR regimes are obtained in \cite{Daut:GLOBECOM:94, Aksu:TCOM:96, Lin:TIT:07} by introducing more advanced schemes at the expense of increased computational complexity.

\subsection{JSCC Code Designs for Realistic Sources}\label{ss:image_practical_code}

\highlight{
In this second class, we can consider JSCC designs that combine compression algorithms for particular source types, e.g., audio, image or video, with channel codes, where the parameters of the two are jointly optimized to maximize the end-to-end performance. In \cite{Modestino:TCOM:79}, the authors combined two-dimensional (2D) differential pulse coded modulation (DPCM) for image coding with a convolutional code. This was later extended to a 2D discrete cosine transform (DCT) in \cite{Modestino:TCOM:79}. Instead, in \cite{Comstock:TCOM:84}, Comstock and Gibson combined 2D-DCT with Hamming codes. Around the same time, there was also significant interest in JSCC for speech signals. Goodman and Sundberg studied the combined effects of quantization and transmission errors on the performance of embedded DPCM in \cite{Goodman:Bell:83a}. Later, they applied these ideas to embedded DPCM speech encoding and punctured
convolutional codes in \cite{Goodman:Bell:83b}. The authors of \cite{Moore:TCOM:84} combined DPCM speech encoding with self-orthogonal convolutional codes.
}

In \cite{peng2000turbo}, a JSCC approach is proposed for Turbo codes to further improve the error control performance of image and video transmission. In this approach, three feedback schemes are introduced to improve the Turbo decoding performance, including error-free source information feedback, error-detected source information feedback, and channel soft values (CSVs) feedback for source signal post-processing. These feedback schemes are implemented by modifying the extrinsic information passed between the constituent maximum a posteriori probability (MAP) decoders in the Turbo decoder. 
Experimental results show that around 
0.6 dB of channel SNR reduction can be achieved by the proposed JSCC schemes without introducing any extra computational cost.

In \cite{wu2003efficient}, 
the authors consider the transmission of  encoded images over noisy channels. 
To provide unequal protection to different bit streams of the image to combat channel noise, the authors consider two rate compatible channel codecs, rate  compatible punctured convolutional codes (RCPC) and rate compatible punctured turbo codes (RCPT), to construct a JSCC system.  In order to derive the optimal rate distribution, they formulate the optimization problem of minimizing the end-to-end distortion, which is then solved by a dynamic programming-based approach.  Experimental results show that the proposed JSCC schemes perform better than the baseline without rate allocation.

The authors of \cite{pu2007joint} consider the rate allocation problem for the transmission of scalable images over parallel channels. 
A JSCC system consisting of an image source codec and a RCPT channel codec is proposed. 
Different from the aforementioned formulation of the optimization problem, distortion-based rate allocation problem for parallel channels is generally difficult to solve, since the search space can be expanded largely and the additional constraints for different subchannel arise.
To tackle this problem, the authors consider a rate-based optimization problem that maximizes the expected length of correctly received data. 
The rate-optimal solution provides a good approximation to the distortion-optimal solution.
Simulation results of the comparison on MSE and PSNR for the transmission of images over parallel Gaussian channels show that the proposed JSCC schemes outperform the traditional approaches.

Different from the above solutions, the authors of \cite{zhang2002power} propose a JSCC scheme with power-minimized rate allocation for video transmission over wireless channel.
The optimization problem is formulated as the minimization of the summation of processing power for source and channel encoder and the transmission power with the constraints of total rate and end-to-end distortion.
Simulation results show that the proposed joint power-control and rate allocation scheme achieves higher power savings compared to the conventional JSCC scheme.

\subsection{Joint Designs}\label{ss:JointDesign}

While the aforementioned designs are considered under the JSCC context, they inherently rely on separate source and channel codes that are either jointly decoded or optimized jointly. On the other hand, there have also been many efforts in the literature to provide genuinely joint designs. The linear coding scheme of Hellman \cite{Hellman:TIT:75}, mentioned previously, can be considered as an example of a joint design, although it is obtained by a direct combination of two linear codes for source and channel coding, respectively.


\highlight{In the case of ideal Gaussian sources and channels, the optimality of uncoded transmission has been well known, \highlight{as highlighted in Section \ref{ss:fundamentals}}. However, this optimality breaks down when there is a mismatch between the source and channel bandwidths, and mapping a multi-dimensional source directly to a multi-dimensional channel has been a challenging open problem for many decades. A common approach is based on the use of space-filling curves for bandwidth compression, originally proposed by Shannon \cite{shannon1949communication} and also studied in depth by Kotelnikov in \cite{kotelnikov}. More recent works include \cite{ramstad2002shannon, hekland2005using, hekland2009shannon}. Hybrid digital and analog schemes that combine vector quantization with analog mappings were considered in \cite{Mittal:IT:02, Skoglund:TIT:06} for robustness. The optimal mapping is identified in \cite{lee1976optimal} when both the encoder and decoder are constrained to  be linear.}

\highlight{Inspired by the theoretical optimality of uncoded transmission for certain source and channel statistics, practical JSCC schemes for image and video transmission were proposed in \cite{lervik1997robust} and \cite{fuldseth1997robust}, respectively. This idea was later popularized as SoftCast in \cite{Jakubczak:10} when uncoded transmission is directly applied after JPEG compression.} SoftCast applies a discrete cosine transform (DCT) on the input image, and transmits the DCT coefficients directly over the channel using a dense constellation. Compression is obtained by discarding blocks of DCT coefficients whose energy is below a certain threshold. On the other hand, index of the discarded blocks is sent as meta-data to the receiver for reconstruction - in that sense, SoftCast is a hybrid scheme combining uncoded/analog transmission with digital communication. Since the encoder mapping is linear for the transmitted coefficients, and the coefficients are corrupted by additive noise directly, the resultant  peak signal-to-noise ratio (PSNR) of the reconstruction at the receiver is linearly related to the channel signal-to-noise ratio (CSNR). This resolves the cliff effect problem encountered in SSCC benchmarks. 

Many variations and improved versions of SoftCast have been introduced in the literature \cite{fan2012wavecast, hekland2009shannon}. A theoretical analysis of SoftCast is presented in \cite{xiong2016analysis}, which highlights the importance of a decorrelation transform and the energy modeling of the underlying signals. Wavelet transform is employed in \cite{fan2012wavecast} and Karhunen-Loeve transform (KLT) is applied in \cite{hagag2017hypercast} as the decorrelation transform instead of 2D-DCT. ECast \cite{zhang2015ecast} focuses on joint sub-carrier matching and power allocation for optimal end-to-end performance of transmission. In \cite{Li:WCNC:11, Schenkel:SPIE:10}, compressed sensing (CS) has been used for wireless video transmission, where $l_1$ approximation is considered for recovery. While this allows the receiver to employ convex optimisation tools, it is still complex for video streaming applications with strict delay constraints. Iterative algorithms have been proposed to approximate the solution faster, which achieve reconstruction through iterative thresholding, at the cost of increased error \cite{Yin:MNA:16}. This algorithm applies CS on the pixels directly and requires no meta-data as long as the measurement matrix is agreed a priori between the transmitter and receiver. 
In \cite{Tung:CL:18}, after applying 2D-DCT on image blocks and thresholding, a novel grouping of the coefficients is applied, where coefficients of the same frequency component are grouped into vectors. Each vector is then multiplied with a pseudo-random measurement matrix, whose size depends on the sparsity level of the corresponding vector. Finally, a scaling factor is applied to the results of this multiplication, which corresponds to power allocation across different frequency components. The receiver employs a combination of approximate message passing (AMP) and minimum mean squared error (MMSE) estimation. AMP is a low-complexity iterative thresholding algorithm, which does not need to know the exact positions of the nonzero elements in the sparse vector. 

\section{Machine learning based JSCC Code Design: Deep Learning Era}\label{s:codes_ml}

As outlined above, there have been ongoing efforts on the design of practical joint source-channel codes over the decades since Shannon, they have not found applications in practical systems; and almost all existing communication systems rely on separate digital codes. While this is partially due to the modularity of separation that makes it attractive from an engineering design point of view, it is also due to the fact that the existing JSCC designs either did not reach the same level of performance achieved by state-of-the-art separate codes, or relied on unwieldy and complex designs that made their implementation in practice infeasible. 

On the other hand, there have been significant progress in practical joint source-channel code design over the past several years thanks to the developments in deep learning technology. 
In general, a joint source-channel code is a pair of mappings as in Section \ref{ss:fundamentals}: the first is the encoder mapping from the source signal space to the channel input space under the channel input cost constraint, while the second is the decoding mapping from the channel output space to the source reconstruction space. Separate source and channel codes, or most JSCC code designs that rely on the adoption and combination of existing hand-crafted source and channel codes, impose certain limitations on the transformations that can be used. In particular, they require mapping of the source signal into a sequence of bits. Such structured codes are more amenable to optimization; however, they are combinatorial problems in essence. This combinatorial nature makes the employment of deep learning techniques in the design  of source and channel codes challenging. In the case of source coding, typically a quantization step must be employed, which is not differentiable, and has to be approximated through various methods (e.g., straight through estimator, stochastic quantization, etc.) \cite{gholami2021survey}. The channel coding problem is also highly challenging as the number of classes grows exponentially with the code length, and the target error probability levels are extremely low compared to typical machine learning tasks, which increases the training time and sample complexity. On the other hand, one can argue that deep learning is a better match for the design of joint source-channel codes, which in essence do not impose any combinatorial constraints, and simply requires mapping similar source signals to nearby points in the channel input space in order to limit the reconstruction distortion in the presence of channel noise. 
In the remainder of this section, we will provide an overview of the recent developments in deep learning aided design of joint source-channel codes, and sketch some of the challenges in their adoption in practical communication systems.  

\begin{figure}[t]
	\centering
	\includegraphics[scale=0.3]{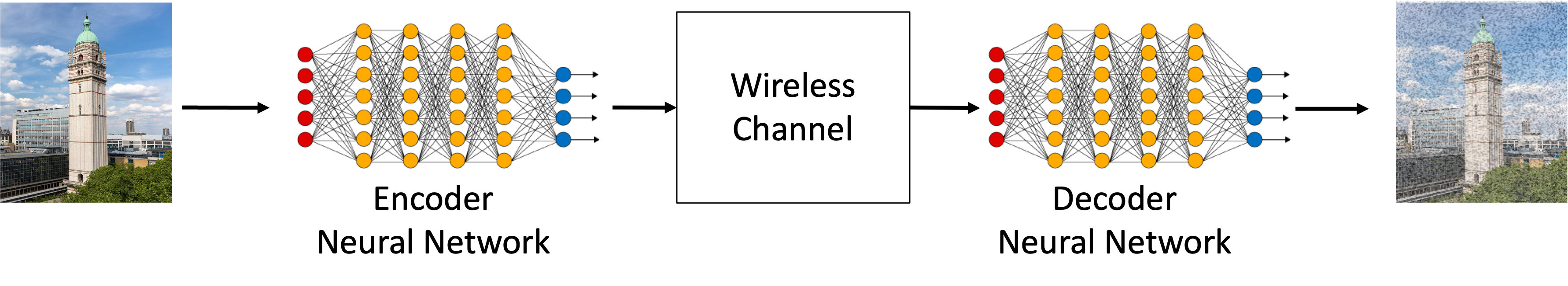}
	\caption{DeepJSCC schemes rely on an autoencoder pair trained jointly on a dataset of input signals and a channel model.}
	\label{fig:DeepJSCC_model}
\end{figure}

\subsection{DeepJSCC for Wireless Image Transmission}\label{ss:image}

Deep learning enabled JSCC was first proposed in \cite{Eirina:TCCN:19} and applied to wireless image transmission. The scheme, called \emph{DeepJSCC}, uses a trainable autoencoder architecture, with a non-trainable channel between the encoder and decoder, to learn a mapping directly from the image pixels to a set of channel input symbols, and vice versa at the decoder (see Fig. \ref{fig:DeepJSCC_model}). \highlight{That is, in DeepJSCC, the encoder and decoder mappings are parameterized by neural networks $\boldsymbol{\Theta}_e$ and $\boldsymbol{\Theta}_d$, and can be denoted by $f^{m,n}_{\boldsymbol{\Theta}_e}$ and $g^{m,n}_{\boldsymbol{\Theta}_d}$, respectively. Here, $m=3 \times h \times w$ denotes the input dimension of the encoder neural network, where $3$ represents the R, G and B components of the colored input image, and $h$ and $w$ denote its height and width, respectively. The corresponding source-channel rate is given by $r=m/n$ as before. In the DeepJSCC literature, the performance is typically quantified in terms of the bandwidth ratio, which is the inverse of the source-channel rate, denoted by $\rho=n/m$.}

\highlight{The two networks are then trained jointly in an end-to-end fashion. The goal is to solve the following optimization problem as before: 
\begin{eqnarray}
    \min_{\boldsymbol{\Theta}_e, \boldsymbol{\Theta}_d} \mathbb{E}\left[d(S^m, \hat{S}^m) \right].
\end{eqnarray}
However, unlike in information theoretic formulation, for practical image communication problem we do not have a statistical model of the source $S^m$, or an additive single-letter distortion measure. Therefore, instead of vying for a closed-form expression, we optimize the neural network parameters using stochastic gradient descent over a dataset of images $\mathcal{D}$:
\begin{eqnarray}
    \min_{\boldsymbol{\Theta}_e, \boldsymbol{\Theta}_d} \frac{1}{|\mathcal{D}|} \sum_{S^m \in \mathcal{D}} \left[d(S^m, \hat{S}^m) \right].
\end{eqnarray}
}

\highlight{We note that this formulation is general, and can be applied to any channel and distortion measure. Here, the channel can be treated as another layer in between the encoder and decoder neural networks, yet an untrainable one. For the standard additive Gaussian noise channel, the gradient trivially backpropagates through the channel layer. In general, it may be difficult to obtain an explicit gradient for the channel model, or we may not even have a channel model, but simply have access to channel input-output pairs. In that case, we can first train a generative model of the channel \cite{Oshea:ICNC:19, Xiao:TWCL:22, Euchner:WSA:24}, which can then be used for training the DeepJSCC encoder/decoder pair.}

\highlight{We would like to highlight that the DeepJSCC approach can be considered as a generalization of the linear/uncoded transmission schemes introduced in Section \ref{ss:JointDesign}, which were limited to hand-designed transformations of the input image. Instead, the solution identified through the training operation can be a highly non-linear mapping from the input source sample to the channel, and in general, its complexity will be dictated by the adopted neural network architecture. Moreover, classical image compression solutions are designed for general natural images, and cannot adapt to the statistics of particular types of images. Instead, DeepJSCC encoder/decoder pair can be trained for a certain source modality or a particular dataset (e.g., face images). On the other hand, for the solution to generalize to arbitrary natural images, we would need a richer training dataset.} 

\highlight{This end-to-end training approach also allows adopting any arbitrary differentiable loss function between the input and its reconstruction, not only the mean-squared error (MSE) loss, i.e., peak signal-to-noise ratio (PSNR). For example, in \cite{Eirina:TCCN:19}, the authors used PSNR as well as the more perceptually aligned structural similarity index measure (SSIM), and showed that DeepJSCC is superior than SSCC under both measures, while the gain under the SSIM loss is even more profound. More recent papers in the literature consider more involved perceptual metrics, such as the learned perceptual image patch similarity (LPIPS) \cite{zhang2018perceptual}, or the FID score \cite{Heusel:NIPS:2017}, which are widely used to quantify the perceptual realism of images \cite{Agustsson:CVPR:23}.}

\begin{figure}[t]
\centering
\includegraphics[width=1\columnwidth]{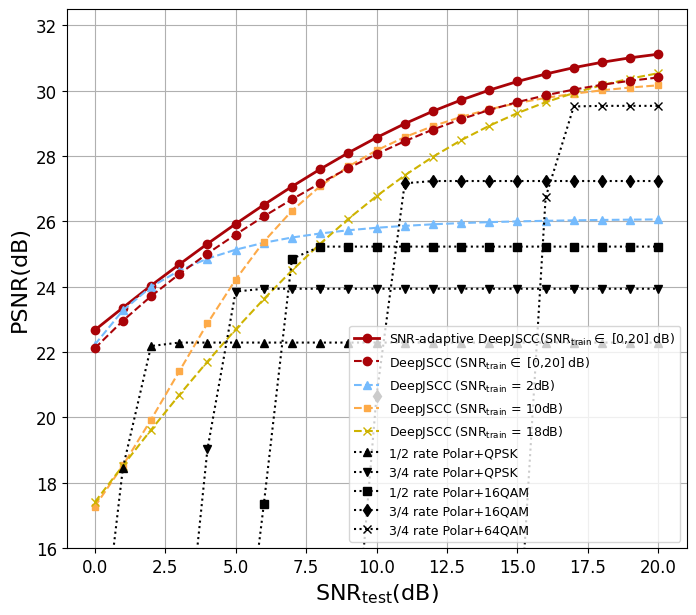}
\caption{Comparison of DeepJSCC performance with that of SSCC on the CIFAR-10 dataset transmitted over an AWGN channel with different SNR values \cite{Xu:ComMag:23}.}
\label{fig:DeepJSCC_performance}
\end{figure}

In Fig. \ref{fig:DeepJSCC_performance}, we present the PSNR performance achieved by DeepJSCC after being trained on a portion of the CIFAR-10 dataset over an additive white Gaussian noise (AWGN) channel in (\ref{AWGN_channel}). Here, each input is a colored image from the dataset of size $32 \times 32$ pixels, and is mapped to a channel input codeword of length $n=256$ symbols. This corresponds to a bandwidth ratio of $\rho = 256/(3 \times 32 \times 32)= 1/12$. Each of the curves denoted by `DeepJSCC' corresponds to an encoder/decoder fully convolutional neural network pair proposed in \cite{Kurka:IZS2020} trained on a different channel signal-to-noise ratio (SNR), $\mathrm{SNR}_{\mathrm{train}}$, denoted by $\boldsymbol{\Theta_{e}}(\mathrm{SNR}_{\mathrm{train}})$ and $\boldsymbol{\Theta_{d}}(\mathrm{SNR}_{\mathrm{train}})$, and tested over a range of channel conditions, $\mathrm{SNR}_{\mathrm{test}} \in [0, 20]~\mathrm{dB}$. Thanks to the fully convolutional architecture, the trained encoder/decoder pair can be used for transmitting images of arbitrary size. The architecture only fixes the source-channel code rate, so that if the input image size is doubled, the transmitted codeword length will also be doubled. 

The performance achieved by a separation-based coding scheme is also included in the figure as the black lines, which employ better portable graphics (BPG) image compression codec \cite{BPG} together with polar codes for channel coding at different rates and constellation sizes. \highlight{Here, we have searched over a range of different code rate and constellation size pairs, and included only the best performing pairs.} As expected, the separation-based scheme has a threshold structure. The code achieves the best performance at a certain $\mathrm{SNR}_{\mathrm{test}}$ value, and its performance drops sharply below this threshold, known as the \textit{`cliff effect'}. Also, the performance remains constant even if the channel quality improves , which is known as the \textit{`levelling-off effect'}. In contrast to the threshold behaviour of the separation scheme, we observe that the DeepJSCC scheme trained for a certain target $\mathrm{SNR}_{\mathrm{test}}$ value exhibits a graceful degradation behaviour; that is, its performance gracefully degrades as the test SNR drops below the training SNR, and it slowly improves when the SNR increases. This behaviour of DeepJSCC is more in line with those of analog modulation schemes, e.g., AM/FM modulation. However, to clarify the distinction, DeepJSCC is still a discrete-time coding scheme, in the sense that, the input signal is converted into discrete samples of potentially continuous real values, which are then mapped to a modulated input waveform (e.g., OFDM) with discrete-time continuous-amplitude parameters. The main difference compared to conventional modulation schemes is the lack of a discrete constellation set (e.g., BPSK or M-QAM), limiting the possible input signals to a finite set of constellation points. 
  
Despite the graceful degradation, we can see from Fig. \ref{fig:DeepJSCC_performance} that the best performance is achieved when the training and test SNRs match. Achieving the convex hull of the performance achieved by all possible such curves would require training and storing a separate pair of neural networks for each $\mathrm{SNR}$ value, or at least for small ranges of $\mathrm{SNR}$ values. This is unlikely to be feasible in practical systems, particularly those that need to operate over a wide range of $\mathrm{SNR}$ values, due to memory constraints. Moreover, it can quickly become intractable when we go beyond a simple single-input single-output (SISO) system considered here.   

One alternative approach is to train a single encoder/decoder pair to be used over a large range of channel conditions. We see in Fig.~\ref{fig:DeepJSCC_performance} that DeepJSCC trained over the whole range of channel SNRs from $0$ to $20$~dB can achieve competitive performance, not significantly below those achieved by encoder-decoder pairs trained at the target SNR values. We note that this scheme is equivalent to communicating without any channel state information (CSI) at either end of communication, while the channel is time-varying. This is equivalent to a block-fading model, where the channel noise variance remains constant during the transmission of each image, but changes independently from one transmission to the next, taking SNR values uniformly distributed between $0$~dB and $20$~dB. This shows that DeepJSCC is capable of learning some form of implicit pilot transmission at the encoder and channel estimation at the decoder, in order to adapt to the varying channel condition. This can be a significant advantage in mobile scenarios when accurate channel estimation can be costly, or even infeasible. 

When the system employs channel estimation and channel state feedback, CSI will be available at both the transmitter and the receiver. In that case, rather than using this information to choose the encoder/decoder parameters from among a set of pretrained networks, one can feed this information to the encoder/decoder architectures, and let the networks pick up the appropriate parameter values according to the channel state. For this, an attention feature (AF) module is introduced in \cite{Xu:TCSVT:22}, which allows adapting the network operations to the CSI, so that in the case of poor channel conditions, only the most important features of the input image are transmitted, but with more protection against noise. The network with the attention feature model is trained with random channel states so that it learns to adapt its parameters to the CSI. The performance achieved by this SNR-adaptive DeepJSCC architecture is the top curve in Fig. \ref{fig:DeepJSCC_performance}, clearly demonstrating that a single network is capable of learning to transmit the input images at any given channel state. 

This idea is further extended to OFDM systems in \cite{Wu:CL:22} through a double attention mechanism. Note that, when there are many parallel channels available for transmission, each with a different quality, the encoder/decoder networks need to learn not only how much error protection to employ over each channel, but also to which channel each input feature must be mapped. Note that, in the case of classical SSCC over parallel channels, we identify the number of bits that can be transmitted reliably over each channel, and the compressed bits are mapped to different channels according to these numbers; that is, more bits are transmitted over the better channel. In the case of the DeepJSCC architecture proposed in \cite{Wu:CL:22}, this is learned through the proposed double attention mechanism. 

Another limitation of the DeepJSCC architectures in \cite{Eirina:TCCN:19, Xu:TCSVT:22, Wu:CL:22} is that a separate network needs to be trained for each source-channel rate. A bandwidth-agile architecture is proposed in \cite{Kurka:TWC:21}, where the image is transmitted in such a way that it can be reconstructed from only a limited portion of the transmitted codeword. This can be considered as a `successive refinement' approach, where the receiver's reconstruction quality increases as it receives more symbols over the channel. A more flexible scheme is considered in \cite{bian2023deepjsccl}, where the available bandwidth is dictated by the higher layers (e.g., the medium access control layer), and is given to the encoder and decoder as part of the CSI. The results in \cite{bian2023deepjsccl} show that a single pair of encoder/decoder networks, built upon the Swin transformer architecture \cite{Liu_2021_ICCV}, can adapt to arbitrary channel SNR and source-channel rate values with only a marginal loss in the performance compared to networks trained for specific channel conditions. This  significantly increases the practicality of DeepJSCC, showing that we only need a single network to be deployed on devices for communication over a large variety of channel conditions. 

\highlight{The idea of JSCC in the finite blocklength regimes relies on mapping the source symbols directly to channel symbols. This not only improves the performance by expanding the set of possible transforms compared to separation-based schemes, which impose a digital bottleneck, but also provides the graceful degradation with noise. When the channel output is a noisy observation of the input signal, the input can be estimated directly from the channel output, the estimation error directly depending on the channel noise. In the case of separation, on the other hand, the receiver first needs to detect the channel codeword, which typically has a threshold behaviour in the case of coding over long blocklengths. However, relying on the estimation of the source signal at the receiver means that the amount of noise in the reconstructed sequence will be random, and depend on the channel realization. This becomes a limitation of DeepJSCC when applied over multi-hop networks, as it results on noise accumulation. In \cite{Zhang:CL:24}, the authors proposed a recursive training method for DeepJSCC, where a separate DeepJSCC encoder/decoder pair is trained specifically for each hop of the network, so that the encoder/decoder pair can adjust to the statistics of the noise images recovered at the relay nodes. This approach improves the performance with respect to using the same encoder/decoder pair throughout the network at the expense of significant increase in complexity; however, it cannot prevent performance degradation as the number of hops increases. An alternative hybrid scheme is proposed in \cite{Bian:ICC:24, Bian:arXiv:24a}, considering the transmission over a network, where only the first hop is wireless, and hence, time-varying, while the remaining hops are through the core network with limited but fixed capacity. Here, DeepJSCC is used only for the first wireless link, while the received noisy channel output is quantized and forwarded digitally through the core network. This scheme mitigates the problem of noise accumulation over multiple hops, while still providing graceful degradation with the quality of the wireless first hop.}

Subsequent works have focused on extending the DeepJSCC architecture in various different directions, including to channels with feedback \cite{Kurka:JSAIT:20, Wu:TWC:24b, zhang2022deep}, to MIMO channels \cite{Wu2023DeepJSCCAdaptiveMIMO, Bian2023SpacetimeMIMO, Wu:TWC:24a}, as well as to multi-user networks \cite{Bian:ICMLCN:24, Yin:ICCC:23, semantic_multihop}. \highlights{The scheme proposed in  \cite{Kurka:JSAIT:20}, called DeepJSCC-f, is inspired by the Schalkwijk-Kailath scheme presented in Section \ref{ss:JSSC_Feedback}. Each image is transmitted over multiple blocks: after each block, the transmitter first recovers the image reconstructed by the decoder using the feedback signal. This reconstruction is then fed into the encoder together with the original image to generate the next channel block. This is shown to outperform the benchmark obtained by BPG compression followed by a capacity-achieving channel code. However, it comes at the cost of increased complexity as a different encoder-decoder pair is trained for each block. More recently, a transformer-based architecture is proposed in \cite{Wu:TWC:24b}, where a single encoder-decoder network pair is trained to refine receiver's reconstruction at each block. This scheme enjoys both improved performance and reduced complexity compared to \cite{Kurka:JSAIT:20}, emphasizing the importance of combining both information theoretic ideas with careful architecture choice when trying to identify the best solutions in practice.}

\highlights{It is shown in \cite{Yilmaz:ICMLCN:24} that DeepJSCC scheme can also benefit from correlated side information at the receiver side, the scenario whose theoretical limits we have presented in Section \ref{ss:SW_JSCC}. In this scenario, the optimality of separation breaks down in the finite blocklength regime, and the scheme presented in \cite{Yilmaz:ICMLCN:24} outperforms the one that combines the state of the art neural network based lossy compression scheme that takes the side information into account \cite{Mital:WACV:23} combined with a capacity-achieving channel code.}

DeepJSCC approaches can also benefit from generative models. A generative model is a type of machine learning model whose goal is to learn the underlying patterns or statistical relations in data in order to generate new unseen samples that are statistically similar to those in the training dataset. Mathematically, the goal of a generative model is to learn a joint distribution of the data features. In the context of JSCC, availability of a generative model for the underlying source data can be considered as a prior knowledge of the source distribution. In a variational autoencoder (VAE), a sample from the latent distribution is mapped to a sample from the data distribution. Therefore, the goal of the transmitter is to convey the right sample from the latent space to the decoder. This idea was used in \cite{Choi:NeurIPS:19} and \cite{Bo:TCOM:24} for JSCC over discrete channels, which are not differentiable, so not convenient for the end-to-end training approach of DeepJSCC. A generative adversarial network (GAN) is used in \cite{Erdemir:JSAC:23}, where the transmitter conveys the Gaussian sample that is then fed into the generator network at the receiver. \highlight{This approach can also be motivated by the information theoretic optimality of uncoded transmission of Gaussian samples. Therefore, one can expect that if a powerful generative model is available at the decoder, the transmitter can be simplified as opposed to the standard DeepJSCC approach. }

\highlight{Generative models can also be used to extend the performance of DeepJSCC to more specific datasets \cite{Erdemir:JSAC:23, Chen:ICASSP:24, Yilmaz:INFOCOM:24}. Consider a transmitter/receiver pair equipped with DeepJSCC encoder/decoder networks trained for a generic image dataset (e.g., ImageNet). Assume that they are then used to transmit a specific type of images, e.g., face images. A generative model of the transmitted images, if available at the receiver, can be used to improve the quality of the image reconstructed by the DeepJSCC decoder, producing much more visually pleasing results especially in extremely low bandwidth and noisy channels. In this context, the generative model can be considered as a side information available at the receiver. In particular, diffusion based generative models have been used in the context of JSCC in \cite{Chen:ICASSP:24, Yilmaz:INFOCOM:24, wu2023cddm}.}

\subsection{Variable-length DeepJSCC}
Constrained by the computational capabilities of a single encoder and decoder, particularly in the case of convolutional networks, the performance of DeepJSCC architecture diminishes as source dimensionality increases, sometimes even underperforming classic separated schemes. One way to improve the performance is to consider a variable-length coding scheme \cite{Yang:ICASSP:22, Zhang:TWC:23}, that is, to dynamically adjust the channel codeword dimension based on the input image. 
An alternative nonlinear transform coding (NTC) based variable-length DeepJSCC scheme is proposed in \cite{Dai:SAC:22}, called \emph{NTSCC}. NTSCC first extracts a semantic latent representation $V \in \mathcal{V}$ of the source signal, and then introduces a conditional entropy model $p_{V | Z}$ in the latent space, i.e.,
\begin{equation}\label{eq_ntscc_entropy_model}
    \begin{aligned}
        p_{V | Z} (V | Z) = & \prod_i \left( \mathcal{N}(v_i|{{\mu}}_i,{{\sigma}}_i^2) * \mathcal{U}(-\frac{1}{2},\frac{1}{2}) \right),
    \end{aligned}
\end{equation}
where a hyperprior $Z$ is learned to serve as side information for approximating the Gaussian distribution parameters ${\mu}_i$ and ${\sigma}_i$ for each latent element $v_i$, and ``$*$'' is a convolutional operation with a standard uniform distribution. The learned entropy model $-\log p_{{ v}_i|Z}({ v}_i|Z)$ indicates the summation of entropy of each $v_i$. Thus, the information density distribution of $V$ is captured.

To achieve adaptive rate allocation for DeepJSCC, the hyperprior $Z$ is regarded as a prior for the DeepJSCC codewords. Specifically, the bandwidth cost, such as the number of OFDM subcarriers, $k_i$ for transmitting $v_i$, can be determined as
\begin{equation}\label{eq_channel_bandwidth_cost_cal}
    k_{i} = Q\Big( -{\eta} \log{p_{v_i|Z}(v_i|Z)} \Big),
\end{equation}
where $Q$ represents scalar quantization, and $\mu$ is a scaling factor. Consequently, with the help of the transformer architecture and a rate attention mechanism, NTSCC closely adapts to the source distribution and facilitates source content-aware transmission. Besides, NTSCC mitigates the disparity between the marginal distribution of latent representations for a specific source sample and the marginal distribution intended for the ensemble of source data for which the transmission model was designed. We note here that the $k_i$ values must be reliably conveyed to the receiver for the receiver to be able to recover the latent representation. In that sense, NTSCC is a hybrid scheme; however, the amount of meta-data that needs to be digitally transmitted is relatively small, and hence, it can be transmitted reliably without sacrificing much of the available channel bandwidth.

Similarly to the rate-distortion modeling in \cite{Balle:NIPS2018}, the optimization problem of NTSCC is formulated within the context of variational modeling, wherein the variational density $q_{{Y, Z} | S}$ is used to approximate the posterior distribution $p_{Y, Z | S}$. This approximation is achieved by minimizing their Kullback-Leibler (KL) divergence over the data distribution $p_{S}$. Accordingly, the optimization of NTSCC system can be formally converted to the minimization of the expected channel bandwidth cost, as well as the expected distortion of the reconstructed data versus the original, which leads to the optimization of the following RD tradeoff,
\begin{equation}\label{eq_expect_loss_func}
    \begin{aligned}
         \mathcal{L}_{\text{BD}} =  \mathbb{E}_{S\sim p_{S}} \Big( \lambda \big( -{\eta} \log{p_{V|Z}(V|Z)} \big) + d(S,\hat{S})\Big),
    \end{aligned}
\end{equation}
where the weight factor $\lambda>0$ dictates the trade-off between the channel bandwidth cost and the distortion. The scaling factor $\eta>0$ correlates the estimated entropy to the channel bandwidth cost and is associated with the source-channel codec capability. 
    
Results in \cite{Dai:SAC:22} demonstrate that the NTSCC method yields significantly improved coding gain and RD performance on well-established perceptual metrics, including PSNR, MS-SSIM, and LPIPS.
Furthermore, achieving the same end-to-end transmission performance, the proposed NTSCC method can lead to great reduction in bandwidth cost when compared to both emerging analog transmission schemes that employ the standard DeepJSCC and classical separation-based digital transmission schemes.

The improved NTSCC architecture in \cite{Wang:STSP:23} introduces a contextual entropy model to capture spatial correlations among semantic latent features more effectively, enabling more precise rate allocation and contextual JSCC. 
It also incorporates an online latent feature editing method to allow for more flexible coding rate control aligned with specific semantic guidance.
Experimental verification demonstrates that the improved NTSCC system achieves approximately a 16.35\% reduction in channel bandwidth compared to the state-of-the-art separation-based baseline combining BPG or VTM image compression with 5G LDPC error correction codes.

Adaptive semantic communication (ASC) system \cite{Dai:SAC:23} emphasizes the practical application of the model based on NTSCC.
Specifically, when confronted with disparities between the distribution of test data or channel responses and the conditions encountered during training, the model may exhibit suboptimal performance. The ASC system leverages the overfitting characteristics of deep learning modules and adapts semantic codecs or representations to individual data or channel state instances. The entire ASC system design is formulated as an optimization problem with the objective of minimizing the loss function, which constitutes a three-way trade-off between data rate, model rate, and distortion terms. While achieving equivalent end-to-end transmission performance based on objective metrics like PSNR, the ASC system can yield up to a 41\% reduction in bandwidth costs compared to the state-of-the-art engineered transmission scheme (VVC combined with 5G LDPC coded transmission).

\highlights{
\subsection{DeepJSCC over Multi-User Channels}\label{ss:multi-user_DeepJSCC}
One of the simplest multi-user network models is the relay channel shown in Fig. \ref{fig:relay_channel}. DeepJSCC over the relay channel has been studied in \cite{Bian:ICMLCN:24, Tang:OJCS:24, bian:arXiv:24b}. A process-and-forward scheme is proposed in \cite{Bian:ICMLCN:24, bian:arXiv:24b}, which relies on DeepJSCC encoding/decoding at the source and destination terminals, as well as a deep learning based processing at the relay. In the case of a half-duplex relay channel, one of the challenges is to decide on the time duration relay listens before starting its transmission. In the case of a decode-and-forward scheme \cite{Cover:TIT:79}, where the relay decodes the message of the source terminal before forwarding it, the relay listen and forward time durations can be adaptive to the source-channel link quality, where the relay listens until it accumulates enough information to guarantee correct decoding of the message. However, this is not possible in the case of DeepJSCC since the relay will always have a noisy estimate of the source signal. This is optimized through a search algorithm in \cite{bian:arXiv:24b}; however, an adaptive mechanism which estimates the quality of relay's estimate can be trained to make this decision adaptively. In the case of a full-duplex relay, on the other hand, the relay can transmit the whole time duration; however, it needs to receive a certain amount of information before starting to forward it. A block coding scheme is proposed in \cite{Bian:arXiv:24a}, inspired by the relaying schemes proposed in \cite{Cover:TIT:79}. In this scheme, the source transmits the information in blocks, and the relay decides on its transmitted signal at each block based on what it has received up to that point. In general, this would require training a separate relay encoder for each block, which learns what to forward in that particular block.  We note that, the relay's goal is not only to forward as much relevant information as possible to the destination, but also to do this in a coherent manner with the source signal, as the two are received at the destination simultaneously. While decreasing the 
number of blocks is beneficial in terms of utilizing the relay (since the relay waits initially until it receives the first block), this would  also increase the number of blocks, and hence, the complexity. In \cite{Bian:arXiv:24a}, a novel transformer based relaying scheme is used throughout all the blocks. This function is trained in a sequence-to-sequence manner, where the previous received signals and transmitted relay codewords are accumulated at each block to increase the amount of relevant information forwarded by the relay. It is also shown that the relay transmits correlated signals with the source, showing that the scheme learns to act as distributed transmit antennas. }

\highlights{In general, the extension of DeepJSCC to broadcast channels is rather trivial. Indeed, transmission over a fading channel without a channel state information available at the transmitter, as studied in \cite{Eirina:TCCN:19}, can also be considered as communicating over a broadcast channel as we can treat each fading channel state as a virtual user with a different channel quality. DeepJSCC over a degraded broadcast channel is studied in \cite{wu2023fusionbased}.}

\highlights{DeepJSCC over a MAC is studied in \cite{Yilmaz:ICC:23}. In the case of a MAC, the challenge is two-fold: on one hand, we would like to reduce the interference among the transmitters. On the other hand, the receiver after recovering the noisy signals should be able to attribute each signal to its correct transmitter. A trivial way to deal with this problem is to train multiple transmitter networks together with the DeepJSCC decoder. However, such a scheme would not scale, as each node will need to have access to all of these encoding functions as it may act as a different encoder at each individual scenario. Instead, the authors in \cite{Yilmaz:ICC:23} propose a scheme in which all the transmitters utilize the exact same DeepJSCC encoder; however, they additionally employ a low-dimensional embedding, which would identify them as a separate transmitter, and allows the receiver to distinguish among different transmitters. This can be considered as a learned code-division multiple access (CDMA) scheme.}

\subsection{DL-aided JSCC for Video Transmission}\label{ss:video}

Video, which accounts for more than 80\% of all internet traffic, is a type of source that is known to be particularly hard to transmit reliably, due to the high data rate requirement to achieve high video quality and frame rate.
Similar to the image case in the previous section, we can design an autoencoder architecture that maps video frames to channel symbols, and vice versa, to learn a JSCC for video transmission.
One key difference between images and videos is the temporal correlation between frames in video.
In a standard video sequence, motion differences between successive frames is typically very small, and video sequences can be represented very efficient by identifying intermittent key frames, which are treated as independent images, and storing only the motion and residual information for the remaining frames, thereby exploiting temporal redundancy. Motion information typically refers to translation vectors, which can be used to translate a given frame to subsequent frames. Since motion vectors cannot capture occlusion and disocclusion of objects, residual information, which captures the difference between the translated frame and the ground truth frame, is also stored. 

Existing works on bringing DeepJSCC to video transmission have largely exploited the aforementioned temporal redundancy among frames to achieve efficient bandwidth utilization.
In \cite{Tung:JSAC:22}, the authors train separate autoencoders to transmit the key frames and temporal information (i.e., motion vectors and residual information) and utilize a reinforcement learning driven policy for allocating channel bandwidth to each frame.
By dynamically allocating channel resources, they show that their scheme, called DeepWiVe, can outperform state-of-the-art video compression codecs H.264 and H.265 when transmitting video over AWGN and Rayleigh fading channels.

In \cite{Wang:JSAC:23}, rather than using reinforcement learning, the authors learn an entropy model of the temporal information and use variable channel uses to dynamically adjust channel resource usage depending on the information content of the frames. 
The idea is that frames with higher entropy require more channel uses and by using variable channel uses, the average channel use can be lower, similar to variable length coding.
They show that their scheme can perform better than H.265 even when assuming capacity achieving channel coding is used.

\subsection{DL-aided JSCC for Text Transmission}\label{ss:text}

As the notion of semantics is traditionally related to natural language, or text in general, semantic communication of text can offer interesting insights into the benefits of DeepJSCC. 
One of the first papers to apply DeepJSCC to text transmission is \cite{Xie:TSP:21}, where the authors proposed DeepSC for text transmission.
DeepSC leverages the transformer architecture, which has been so successful at natural language processing (NLP) tasks, to extract semantic features that are useful for JSCC.
To measure the distortion, the authors propose to use a pretrained large language model, BERT \cite{Peters:NAACL:2018}, and compute the cosine similarity between two sentence in the embedding space of BERT.
The idea is that, having been trained on a large corpus of text data, BERT has an internal representation of natural language that is consistent with semantic meaning.
Therefore, the cosine similarity between the embeddings of two sentences represent the semantic similarity between them.

Similar to what was observed in DeepJSCC for image and video transmission, it is also observed that the performance of DeepSC, as measured by the bilingual evaluation understudy (BLEU) score \cite{Papineni:ACL:2002}, gracefully degrades as the channel noise increases, unlike the cliff-edge drop off that SSCC exhibits.
Moreover, DeepSC performs much better than SSCC under fading channel conditions.
Perhaps more interestingly, DeepSC is more likely to produce coherent sentences that may not be exactly the same as the source, but is semantically similar, showing that not only is the loss function effective at capturing source semantics, it is also a good illustration of how semantic communication is related to JSCC. 

Further works have address other issues with text transmission. For example, in \cite{Yan:LWC:22}, the spectral efficiency problem is addressed by first defining a semantic spectral efficiency metric, that measures on average how much semantic information is transmitted per unit of bandwidth (usually measured in hertz).
The authors define semantic information using the semantic loss used to train DeepSC, and maximize the average spectral efficiency over all users. 
In \cite{Peng:GLOBECOM:2022}, the authors address potential adversarial attacks on the DeepSC transceiver by improving the underlying attention mechanism in the transformer layers.
In essence, they train an additional layer at the output of the attention layer to detect which input tokens are potentially corrupted.
This information is then passed to the subsequent attention layers to inform it of which tokens to ignore.


\subsection{Other Sources and Channels}\label{ss:others}
Naturally, JSCC can be applied to many other source signals, and the application of neural networks for the design of competitive DeepJSCC schemes allowed researchers to design end-to-end communication systems for a wide variety of source and channel distributions and applications.

One of the promising applications of DeepJSCC is for CSI feedback in massive MIMO systems, where the base station requires accurate and timely channel state information (CSI) for beamforming and interference management. In a frequency division duplex (FDD) massive MIMO system, user equipments (UEs) estimate their downlink CSI based on the pilots transmitted by the BS. This CSI estimates then need to be transmitted to the BS over the uplink channel. It is, hence, crucial to design an efficient feedback scheme that provides more accurate CSI to the BS while introducing a limited feedback overhead. There has been significant progress in the recent years in the design of efficient CSI compression techniques using DNNs \cite{Wen:WCL:18, Liao:Access:19, Mashhadi:TWC:21}. However, since these rely on digital compression techniques, they need accurate estimation of the uplink channel state. An alternative analog CSI feedback scheme is proposed in \cite{Mashhadi:ICASSP:20}, which directly maps the estimated channel coefficients to channel input using DeepJSCC. It is shown in \cite{Mashhadi:ICASSP:20} that the joint design can significantly improve the achieved downlink communication rate compared to separation. This approach was later extended in \cite{Xu:JSAC:23} by incorporating SNR adaptivity. 

Another source that can benefit from the JSCC approach is point cloud, which refers to a set of points that collectively depict a physical object or a scene \cite{shao2024point}. A point cloud conveys not only the geometry of the object, through the coordinates of the points in the space, but also various attributes associated with each point, such as color, intensity, curvature, etc. It has found extensive applications across different fields in recent years, including robotics, autonomous driving, VR/AR, and metaverse. While there has been significant research efforts for the design of more efficient compression algorithms for point clouds, recent works show that JSCC can provide gains also for this challenging source modality \cite{Fujihashi:ICC:19, bian2023wireless, Fujihashi:TM:22}. 

With the increasing success and popularity of deep learning models for a wide variety of tasks, it is expected that another type of information source that may constitute a significant portion of future wireless data traffic is model weights. While each day new results appear showing that neural networks achieve state-of-the-art performance in yet another task, it is not reasonable to assume that a mobile device will be capable of storing all possible neural networks locally to be deployed at the time of need. Instead, it is more likely that the model parameters for a desired task will be downloaded on demand from a server when they are needed. This requires transmitting those parameters as quickly and efficiently as possible over wireless channels. In \cite{Jankowski:ISIT:22, Jankowski:TWC:24}, authors proposed a joint training and transmission strategy, where the model parameters are mapped directly to the channel inputs, such that a noisy version of the model will be recovered at the receiver depending on the channel conditions. 

In addition to different source modalities mentioned above (speech transmission is considered in \cite{Weng:JSAC:21, Han:JSAC:23}), researchers have also studied the design of DeepJSCC solutions for different channels, including image transmission over underwater acoustic channels \cite{Anjum:WUWNet:22}, or image storage in memristive devices \cite{Zheng:IEDM:18, Zarcone:SR:20} or even in DNA sequences \cite{wu2023deep}.

\section{Further Discussion and Conclusion} \label{s:conclusion}

In this paper, we  have presented a comprehensive overview of the foundations of JSCC. While separation provides modularity for the design of complex communication networks, it is known to be suboptimal in 
practical scenarios when the blocklengths are finite. Yet, after decades of research, state-of-the-art SSCC schemes present a benchmark that is hard to beat, particularly when the blocklength is relatively large. On the other hand, in the case of non-ergodic communication scenarios, or when the blocklength is extremely limited due to resource and latency constraints, JSCC becomes an attractive alternative. Moreover, in the case of many multi-user communication scenarios, we do not even have theoretical optimality of separation, and the optimal JSCC scheme is not known for even very simple multiple-access and broadcast channel models. After a brief review of classical approaches to practical design of JSCC schemes, we have mainly focused on recent developments in JSCC design benefiting from deep learning techniques for various source and channel combinations. It has been recently shown that DeepJSCC approaches can outperform state-of-the-art separate designs, making them a very attractive alternative, particularly for mobile communication scenarios where channel state changes rapidly over time, and its accurate estimation would require significant channel resources. In the light of these significant developments in DeepJSCC design over the past several years, we believe it is time to start questioning the strict layered architecture of contemporary communication systems, and explore alternative joint approaches. 

\highlight{Despite the fact that almost all current communication systems rely on the SSCC principle, there are many potential immediate applications of JSCC in standalone point-to-point communication systems. An existing successful example is an implementation of JSCC for MIMO systems by Amimon, based on unequal error protection. This technology was adopted in professional wireless cameras that were used by many Hollywood production companies, and won an Academy Award for Technical Achievement from the Academy of Motion Picture Arts and Sciences in 2021.} 

\highlight{Another potential application of JSCC is the transmission of video signals from a drone to a ground station. Current technology relies on WiFi, and results in the loss of video signal beyond few kilometers, particularly in busy urban areas. Adoption of JSCC techniques, e.g., DeepJSCC, for such a point-to-point link can enable constant connectivity, where the quality of received video gracefully degrades depending on the channel condition, but is never completely lost. Moreover, it can also work without channel estimation, which is difficult and costly to obtain in highly mobile scenarios. Another application is AR/VR headsets, which require the transmission of high-rate multimedia content. Most current headsets rely on wired connections or very high-quality stable wireless links, which considerably limits their functionality. }

\highlight{On the other hand, despite their superiority from a purely performance point of view, there are still many practical and fundamental challenges that need to be addressed for the potential adoption of JSCC schemes in 6G or future communication standards. First of all, a full-fledged implementation of JSCC requires significant changes in the network architecture as it relies on the joint optimization of the physical layer coding and modulation together with source compression, which is normally handled at the application layer. This creates a fundamental problem in practical applications in mobile networks beyond architectural changes, as normally the content to be transmitted belongs to the user, and is not accessible by the network provider. Hence, JSCC needs to be implemented without having access to the underlying content at the encoder. Alternatively, the content providers can apply JSCC directly and convey the coded information to the lower layers, which informs the transmitter what to transmit without seeing the content. This can be done by using DeepJSCC-Q as proposed in \cite{Tung:JSAC:22}, which implements DeepJSCC on a finite constellation. However, to achieve the potential benefits of DeepJSCC requires using a large constellation (to sufficiently approximate continuous-amplitude transmission), and this would require a large bandwidth to convey such a code over the core network until it reaches the wireless link. }

\highlight{Another challenge in implementing JSCC in legacy network architectures is the adoption of feedback, in particular how to benefit from automatic repeat request (ARQ) tools that are widely used in current network protocols. In the current framework, since the goal is to recover a channel codebook reliably, a cyclic redundancy check (CRC) is added to the data packet to verify correct decoding, and if the transmission fails, the receiver sends a negative acknowledgement (NACK) signal requesting retransmission. This is repeated until the data packet is correctly decoded, or the maximum number of retransmissions is reached. On the other hand, in the case of JSCC, typically a noisy version of the source data is recovered at the receiver, whose quality depends on the channel quality. Therefore, it is not possible to make a binary decision at the receiver. One alternative could be to make a decision whether a desired quality level is reached or not, but it is also difficult to make such a decision solely based on the reconstructed signal at the receiver, without having access to the original source signal. This may require rethinking the whole feedback system in future communication networks going beyond the current single-bit ARQ framework \cite{Wu:TWC:24b}.}

\highlight{Adoption of DeepJSCC in standards would also require it to be back-compatible, and allow its implementation on legacy terminals. Many existing devices use chipsets which already implement current codes and modulation schemes on their hardware, which may be difficult to circumvent, or require JSCC solutions around these existing codes. Another critical challenge is the security. JSCC schemes mostly rely on the correlation between the transmitted signal and the underlying source signal. While this results in graceful degradation in reconstruction quality at the legitimate receiver as the channel quality drops, it also means that a signal received by a potential eavesdropper in the vicinity will leak information about the source signal. Digital transmission allows applying encryption techniques where the transmitted bits can be completely independent of the source bits, e.g., through a one-time pad, whereas it is not clear how to achieve such an encryption on the continuous amplitude source signal. While some initial solutions to the security problem in DeepJSCC have been proposed \cite{Marchioro:ICASSP:20, Tung:ICC:23, Kalkhoran:Globecom:23, Chen:Globecom:23, Nan:JSAC:23}, more significant research is needed before they meet the necessary security guarantees for implementation in mobile networks or other systems communicating sensitive content.}

\bibliographystyle{IEEEtran}
\bibliography{references.bib}

\end{document}